\documentclass[twocolumn]{aastex62}

\makeatletter

\newcommand{\Rmnum}[1]{\expandafter\@slowromancap\romannumeral #1@}
\makeatother
\shortauthors{Rui Wang  et al.}

\begin{document}

\title{Properties of Radial Velocities measurement based on LAMOST-\Rmnum{2} Medium-Resolution Spectroscopic Observations}

\correspondingauthor{A-Li Luo and Jian-Jun Chen}
\email{lal@nao.cas.cn; jjchen@nao.cas.cn}

\author{R. Wang}
\affil{Key Laboratory of Optical Astronomy, National Astronomical Observatories, Chinese Academy of Sciences, Beijing 100101, China}
\affil{University of Chinese Academy of Sciences, Beijing 100049, China}

\author[0000-0002-0786-7307]{A.-L. Luo}
\affil{Key Laboratory of Optical Astronomy, National Astronomical Observatories, Chinese Academy of Sciences, Beijing 100101, China}
\affil{University of Chinese Academy of Sciences, Beijing 100049, China}

\author{J.-J. Chen}
\affil{Key Laboratory of Optical Astronomy, National Astronomical Observatories, Chinese Academy of Sciences, Beijing 100101, China}

\author{Z.-R.Bai}
\affil{Key Laboratory of Optical Astronomy, National Astronomical Observatories, Chinese Academy of Sciences, Beijing 100101, China}

\author{L. Chen} 
\affil{Key Laboratory for Research in Galaxies and Cosmology, Shanghai Astronomical Observatory, Chinese Academy of Sciences, 80 Nandan Road, Shanghai 200030, China}
\affil{University of Chinese Academy of Sciences, Beijing 100049, China}

\author{X.-F. Chen}
\affil{Key Laboratory for the Structure and Evolution of Celestial Objects, Yunnan Observatories, Chinese Academy of Sciences, Kunming 650216, China}

\author{S.-B. Dong}
\affil{Kavli Institute for Astronomy and Astrophysics, Peking University, Beijing 100871, China}

\author{B. Du}
\affil{Key Laboratory of Optical Astronomy, National Astronomical Observatories, Chinese Academy of Sciences,Beijing 100101, China}

\author{J.-N. Fu}
\affil{Department of Astronomy, Beijing Normal University, Beijing 100875, China}

\author{Z.-W. Han}
\affil{Key Laboratory for the Structure and Evolution of Celestial Objects, Yunnan Observatories, Chinese Academy of Sciences, Kunming 650216, China}

\author{J.-L. Hou} 
\affil{Key Laboratory for Research in Galaxies and Cosmology, Shanghai Astronomical Observatory, Chinese Academy of Sciences, 80 Nandan Road, Shanghai 200030, China}
\affil{University of Chinese Academy of Sciences, Beijing 100049, China}

\author{Y.-H. Hou }
\affil{Nanjing Institute of Astronomical Optics, \& Technology, National Astronomical Observatories, Chinese Academy of Sciences, Nanjing 210042,  China}
\affil{University of Chinese Academy of Sciences, Beijing 100049, China}

\author{W. Hou}
\affil{Key Laboratory of Optical Astronomy, National Astronomical Observatories, Chinese Academy of Sciences, Beijing 100101, China}

\author{D.-K. Jiang}
\affil{Key Laboratory for the Structure and Evolution of Celestial Objects, Yunnan Observatories, Chinese Academy of Sciences, Kunming 650216, China}

\author{X. Kong}
\affil{Key Laboratory of Optical Astronomy, National Astronomical Observatories, Chinese Academy of Sciences, Beijing 100101, China}

\author{L.-F. Li}
\affil{Key Laboratory for the Structure and Evolution of Celestial Objects, Yunnan Observatories, Chinese Academy of Sciences, Kunming 650216, China}

\author{C. Liu}
\affil{Key Laboratory of Optical Astronomy, National Astronomical Observatories, Chinese Academy of Sciences, Beijing 100101, China}
\affil{University of Chinese Academy of Sciences, Beijing 100049, China}

\author{J.-M. Liu}
\affil{Key Laboratory of Optical Astronomy, National Astronomical Observatories, Chinese Academy of Sciences,Beijing 100101, China}

\author{L. Qin}
\affil{Key Laboratory of Optical Astronomy, National Astronomical Observatories, Chinese Academy of Sciences, Beijing 100101, China}
\affil{University of Chinese Academy of Sciences, Beijing 100049, China}

\author{J.-R. Shi}
\affil{Key Laboratory of Optical Astronomy, National Astronomical Observatories, Chinese Academy of Sciences, Beijing 100101, China}
\affil{University of Chinese Academy of Sciences, Beijing 100049, China}

\author{H. Tian}
\affil{Key Laboratory of Optical Astronomy, National Astronomical Observatories, Chinese Academy of Sciences, Beijing 100101, China}

\author{H. Wu}
\affil{Key Laboratory of Optical Astronomy, National Astronomical Observatories, Chinese Academy of Sciences, Beijing 100101, China}
\affil{University of Chinese Academy of Sciences, Beijing 100049, China}

\author{C.-J. Wu}
\affil{Key Laboratory of Optical Astronomy, National Astronomical Observatories, Chinese Academy of Sciences, Beijing 100101, China}

\author{J.-W. Xie}
\affil{School of Astronomy and Space Science, Nanjing University, Nanjing 210093, China; Key Laboratory of Modern Astronomy and Astrophysics in Ministry of Education, Nanjing University, Nanjing 210093, China}

\author{H.-T. Zhang}
\affil{Key Laboratory of Optical Astronomy, National Astronomical Observatories, Chinese Academy of Sciences, Beijing 100101, China}

\author{S. Zhang}
\affil{Key Laboratory of Optical Astronomy, National Astronomical Observatories, Chinese Academy of Sciences, Beijing 100101, China}
\affil{University of Chinese Academy of Sciences, Beijing 100049, China}

\author{G. Zhao}
\affil{Key Laboratory of Optical Astronomy, National Astronomical Observatories, Chinese Academy of Sciences, Beijing 100101, China}
\affil{University of Chinese Academy of Sciences, Beijing 100049, China}

\author{Y.-H. Zhao}
\affil{Key Laboratory of Optical Astronomy, National Astronomical Observatories, Chinese Academy of Sciences, Beijing 100101, China}
\affil{University of Chinese Academy of Sciences, Beijing 100049, China}

\author{J. Zhong} 
\affil{Key Laboratory for Research in Galaxies and Cosmology, Shanghai Astronomical Observatory, Chinese Academy of Sciences, 80 Nandan Road, Shanghai 200030, China}

\author{W.-K. Zong}
\affil{Department of Astronomy, Beijing Normal University, Beijing 100875, China}

\author{F. Zuo}
\affil{Key Laboratory of Optical Astronomy, National Astronomical Observatories, Chinese Academy of Sciences, Beijing 100101, China}
\affil{University of Chinese Academy of Sciences, Beijing 100049, China}

\begin{abstract}
The radial velocity (RV) is a basic physical quantity which can be determined through Doppler shift of the spectrum of a star.  The precision of  RV measurement depends on the resolution of the spectrum we used and the accuracy  of wavelength calibration. In this work, radial velocities of LAMOST-II medium resolution (R $\sim$ 7500) spectra are measured for 1,594,956 spectra (each spectrum has two wavebands) through matching with templates. A set of RV standard stars are used to recalibrate the zero point of the measurement, and some reference sets with RVs derived from medium/high-resolution observations are used to evaluate the accuracy of the measurement.  Comparing with reference sets, the accuracy of our measurement can get 0.0227 $\text{km}~\text{s}^{-1}$ with respect to radial velocities standard stars. The intrinsic precision is estimated with the multiple observations of single stars, which can achieve to 1.36 $\text{km}~\text{s}^{-1}$,1.08 $\text{km}~\text{s}^{-1}$, 0.91 $\text{km}~\text{s}^{-1}$ for the spectra at signal-to-noise levels of 10, 20, 50, respectively. 
\end{abstract}


\section{Introduction}           
\label{sect:intro}

The radial velocity (RV) is a basic and key physical quantity in the kinematic and dynamic study of our Galaxy. RVs of stars can be derived from the Doppler shifts of their spectra. Many spectroscopic sky surveys have released large samples of RVs, for example SDSS/SEGUE~\citep{2009AJ....137.4377Y}, SDSS/APOGEE~\citep{2017AJ....154...94M,2015AJ....150..148H,2018AJ....156..125H}, LAMOST/LEGUE~\citep{2012RAA....12.1197C,2015RAA....15.1095L,2012RAA....12..735D,2012RAA....12..723Z}, RAVE~\citep{2006AJ....132.1645S,2017AJ....153...75K}, Gaia-ESO Survey~\citep{2012Msngr.147...25G}, Gaia-RVS~\citep{2004MNRAS.354.1223K,2018A&A...616A...5C}, HERMES-GALAH~\citep{2015MNRAS.449.2604D}, etc. These big samples of RVs were obtained through spectra with different resolving power, which provides astronomers useful tools for dissecting and understanding the structure of the Milky Way.

A large sample with consistent measured RVs is a key underpinning for a lot of research work, for example,  \cite{2015AJ....150...97G} use RVs to determine membership of a stellar cluster. Besides, researchers analyzed the RV variations to constrain the stellar pulsation model~\citep{2018MNRAS.474.3344B} and to determinate the properties of the eclipsing binary stars~\citep{ 2018arXiv181204319H,2019arXiv190101627M}.  

LAMOST started a medium-resolution survey (MRS: Liu Chao et al. 2019. in preparation) after its first 5-year low-resolution survey~\citep{2012RAA....12.1197C,2015RAA....15.1095L,2012RAA....12..735D,2012RAA....12..723Z}, the new MRS is driven by several scientific motivations (Galactic archeology, time-domain astronomy, star formation etc, details in Liu Chao et al. 2019. in preparation). Beside, contemporary and future space-based projects (such as Gaia, Transiting Exoplanet Survey Satellite~\citep[TESS;][]{2014SPIE.9143E..20R}, etc.) providing new scientific prospect, LAMOST MRS also develop the corresponding observation plan in Gaia field, TESS field and Kepler field~\citep{2019arXiv190100619L} for producing valuable information from medium resolution spectra including radial velocities and stellar parameters (Liu Chao et al. 2019. in preparation).

The performance of RV measurement is the basis to realize the science goal of the LAMOST MRS survey. The precision of RV measurements from stellar spectra is limited by many aspects, such as,  spectral  resolution, wavelength caiibration, spectral type, spectral range, and the measurement methods~\citep{2001A&A...374..733B}. In this work, we try to make use of the whole LAMOST MRS spectral information to measure RVs for all type of stars. Also, the MRS spectra of stars with repeat observations are used to estimate intrinsic precision of the RV measurement for MRS spectra, while some standard stars are used to derive the accuracy of the RV measurement. The paper is organized as follows. LAMOST-\Rmnum{2} medium-resolution spectroscopic observation and data reduction are described in Sect.\ref{sect:data}. The methods for the determination of RV are presented in Sect.\ref{sect:method}. Calibration and Validation results are highlighted in Sect.\ref{sect:results}. Finally, we summarize this work in Sect.\ref{sect:summary}.

\section{Data}
\label{sect:data}

\subsection{Observations}
\begin{figure}[htbp]
\centering
\includegraphics[width=0.5\textwidth, angle=0]{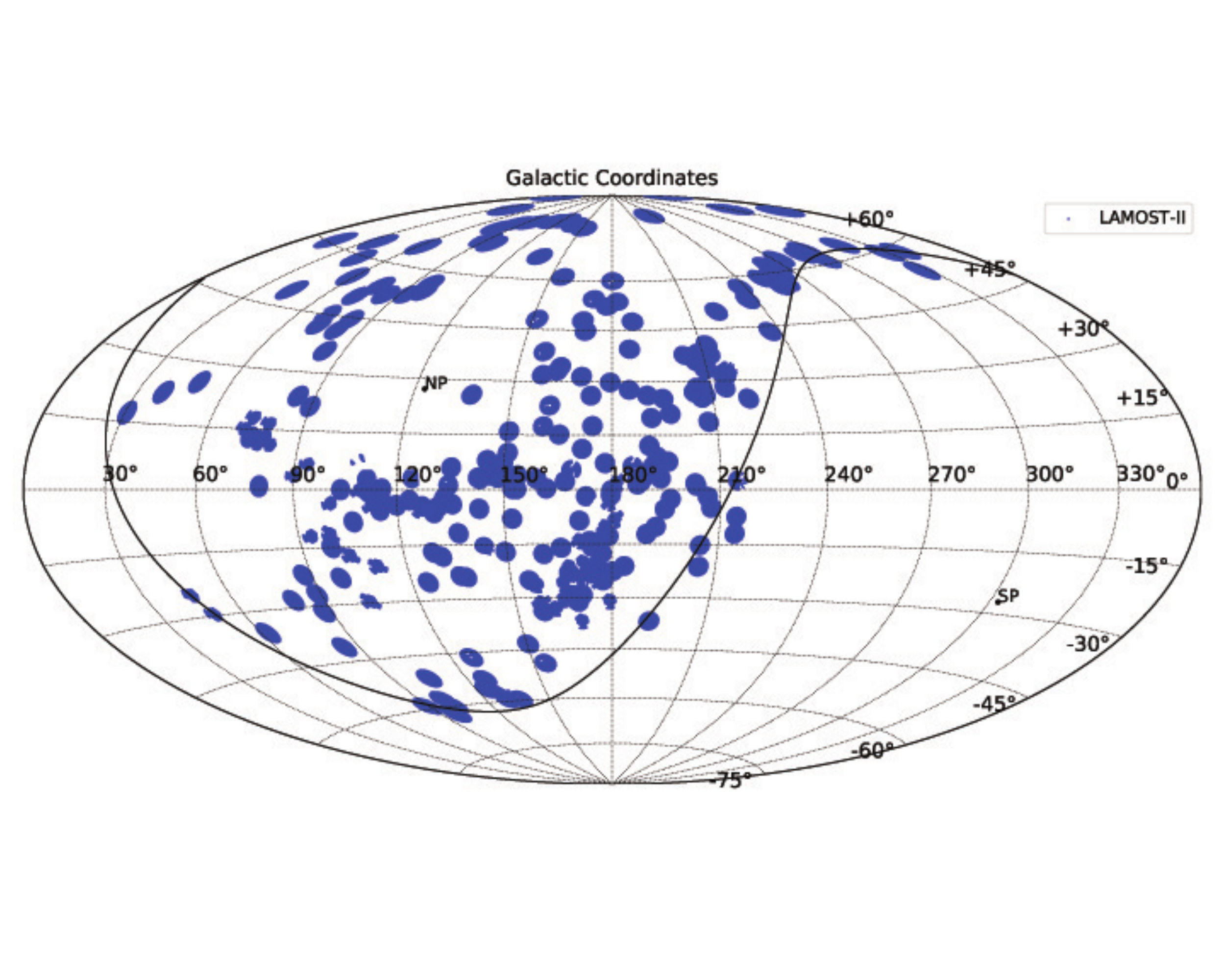}
\caption{The ``footprints" of LAMOST-\Rmnum{2} MRS observations. The projection is in Galactic Coordinates. NP and SP in the figure refer to the North Pole and the South Pole of celestial coordinates. Averagely, spectra in each blue circle are around 3,000 targets. The date of the observation is from 30th Jun 2017 to 31st Dec 2018.}
\label{Figure1}
\end{figure}

\begin{figure}[htbp]
\centering
\includegraphics[width=0.5\textwidth, angle=0]{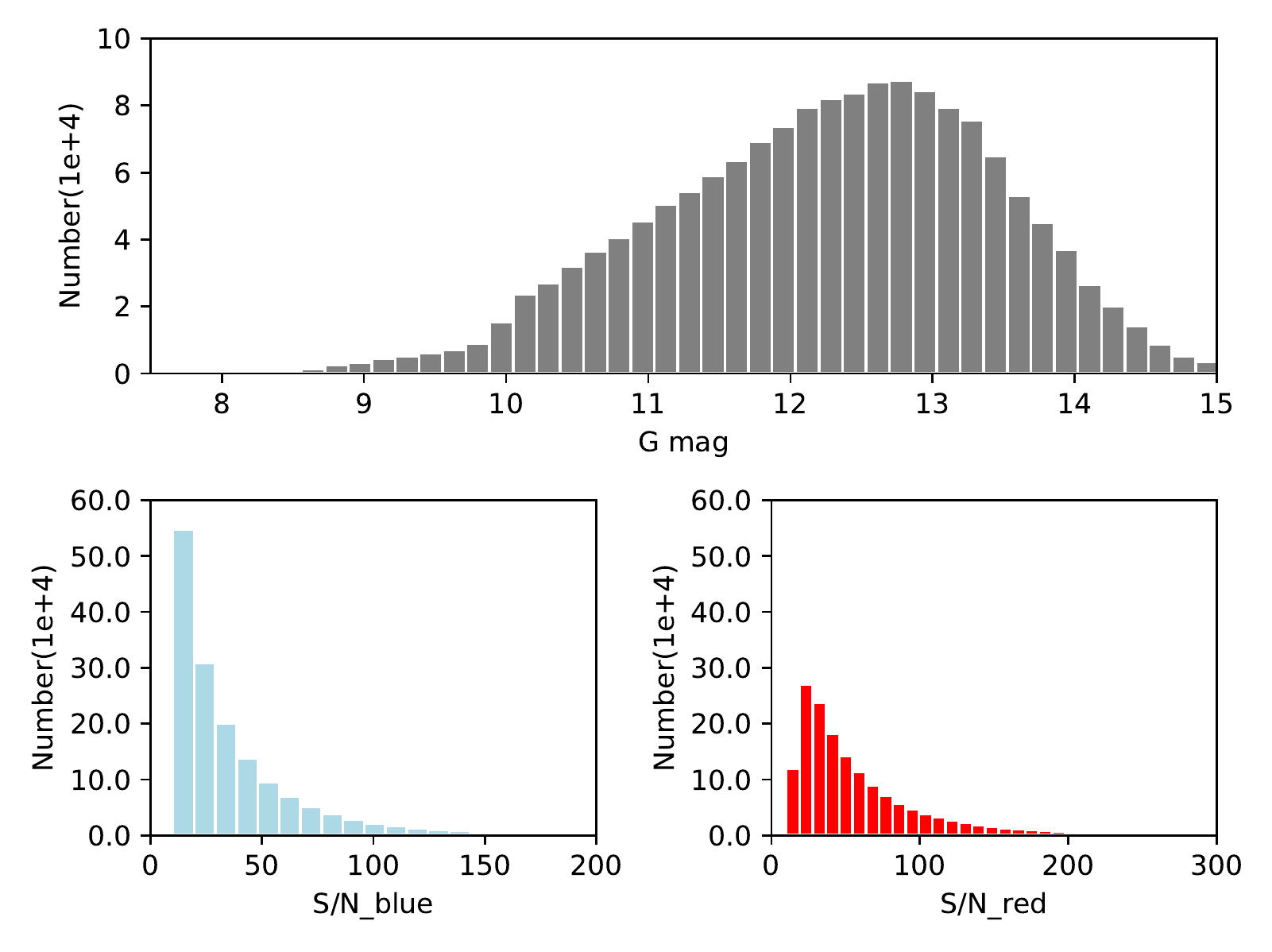}
\caption{Distribution of the G magnitude and signal-to-noise (S/N) of LAMOST-\Rmnum{2} MRS test observations are shown at the top panel and bottom two panels respectively. The date of the observation is from 30th Jun 2017 to 31st Dec 2018.}
\label{Figure2}
\end{figure}

LAMOST is a telescope possessing an effective aperture of 4 meters and $5^\circ$ fields of view, which locates at Xinglong Observatory, Hebei Province, China. The light of 4000 stellar objects that are simultaneously observed is transmitted to 16 spectrometers through 4000 fibers and then recorded by 32 4K X 4K Charge-Coupled Devices (CCD) cameras. LAMOST spectrograph has two resolving modes: the low-resolution mode of R=1800 and the medium-resolution mode of R=7500. After phase one (LAMOST-\Rmnum{1}) low-resolution survey (LRS) from 24th Oct 2011 to 16th Jun  2017~\citep{ 2015RAA....15.1095L}, a period of test observation of the medium-resolution survey (MRS) for LAMOST phase two (LAMOST-\Rmnum{2}) began on the 1st Sep 2017. Until the 31st Dec 2018,  1,597,675 spectra of 281,515 stars from the MRS test observation have been collected, and each spectrum consists of two wavelength bands, i.e. a blue band (4900-5400 \AA) and a red band (6300-6800 \AA). According to wavelength ranges and the size of CCD, each pixel width is 0.12\AA ~(500 \AA /4096 pixels). The ``footprints" of LAMOST-\Rmnum{2} MRS test observations are shown in Fig.\ref{Figure1}. The distribution of the corresponding $\text{G}_{\text{RVS}}$ magnitude adopted from the Gaia DR2 photometer catalogue and the distribution of signal-to-noise (S/N) of the test observations are respectively shown in Fig.\ref{Figure2}. Hereafter, S/N is defined as an average value in a wavelength band and indicates the S/N \emph{per pixel}.

\subsection{Data Reduction}
\begin{figure*}[htbp]
\centering
\includegraphics[width=\textwidth, angle=0]{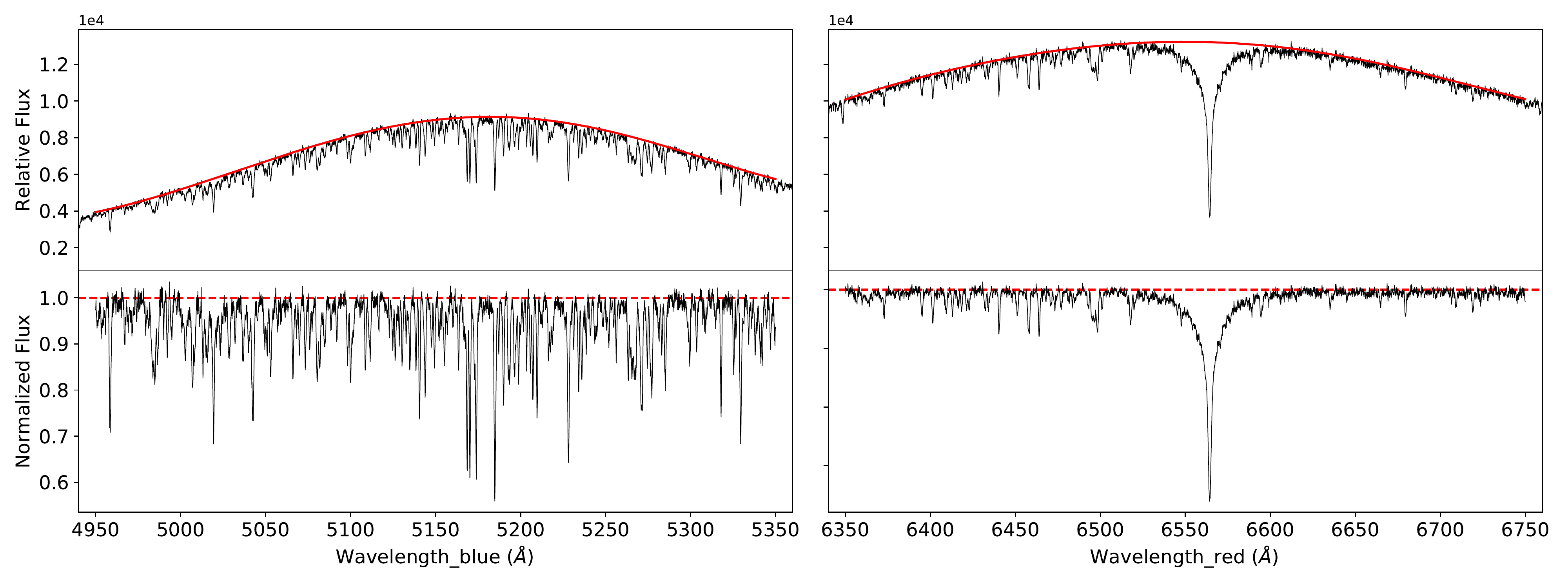}
\caption{An example of a LAMOST MRS spectrum: relative flux (top) and continua normalized flux (bottom) for blue (left) and red (right) waveband. The red solid curves in the top panels are the pseudo-continuum and the red dotted lines in the bottom panels are plotted as a reference.}
\label{Figure3}
\end{figure*}

The MRS spectra are extracted from raw data (CCD images) with the LAMOST reduction pipeline, which is the same processing as that of low-resolution spectra~\citep{2015RAA....15.1095L}. Different from using Cd-Hg and Ar-Ne lamps to carry out wavelength calibration for low-resolution spectra, the dispersion curves derived from Th-Ar and Sc lamps are used for the wavelength calibration of MRS spectra. Appropriate barycentric corrections were applied during the wavelength calibration process. The flux of each spectrum is rectified for both blue and red band, which means the continua are normalized to ``one".  To eliminate the effect of strong emission lines during the calculation of RV, we especially mask the normalized flux higher than 10 percent of the continuum. For the spectra with S/N higher than 10, basic stellar parameters have been calculated using the LAMOST Stellar Parameter pipeline~\citep[LASP;][]{2015RAA....15.1095L, 2011RAA....11..924W}. An example of MRS spectra with two wavebands of a star is shown in Fig.\ref{Figure3}, which include both before and after the continua normalized. 

\section{Methods}
\label{sect:method}

The basic idea of RV measurement is to pick out the largest peak from a group of correlation functions, which are calculated between an observed spectrum and each synthetic template of the 2,194 Kurucz  spectra~\citep{2004astro.ph..5087C}. This full template fitting uses all pixels in all spectral lines rather than a few line centers to reduce the error of wavelength calibration caused by the uncertainty of the dispersion function of gratings. The stellar parameter coverage of the synthetic template grids is shown in Table.\ref{Tab:params_grids}. Both the MRS spectra and synthetic spectra are continuum-normalized using a method the same as \citet{2008AJ....136.2022L}. 

\begin{table}[!h]
\caption{Stellar Parameters Space of the Grid.}
\begin{center}
\label{Tab:params_grids}
\begin{tabular}{lclclcl}
\hline\noalign{\smallskip}
\hline\noalign{\smallskip}
Variable 									&  		Range  							&  Step size  	\\
\hline\noalign{\smallskip}
$\text{T}_{\text{eff}}$ (K)			&  3500 $\sim$ 10000		&  250			\\                        
$\text{log}~\textit{g}$ (dex)	& 	  0.0 $\sim$ +5.0    		&  0.5				\\
$\text{[Fe/H]}$ (dex) 					&    -2.5 $\sim$ +0.5  			&  0.5				\\
\noalign{\smallskip}\hline
\end{tabular}
\tablecomments{ Besides seven [Fe/H] values in the list, there are two additional values of -4.0 and +0.2 dex.}
\end{center}
\end{table}

\begin{figure*}[htbp]
\centering
\includegraphics[width=\textwidth, angle=0]{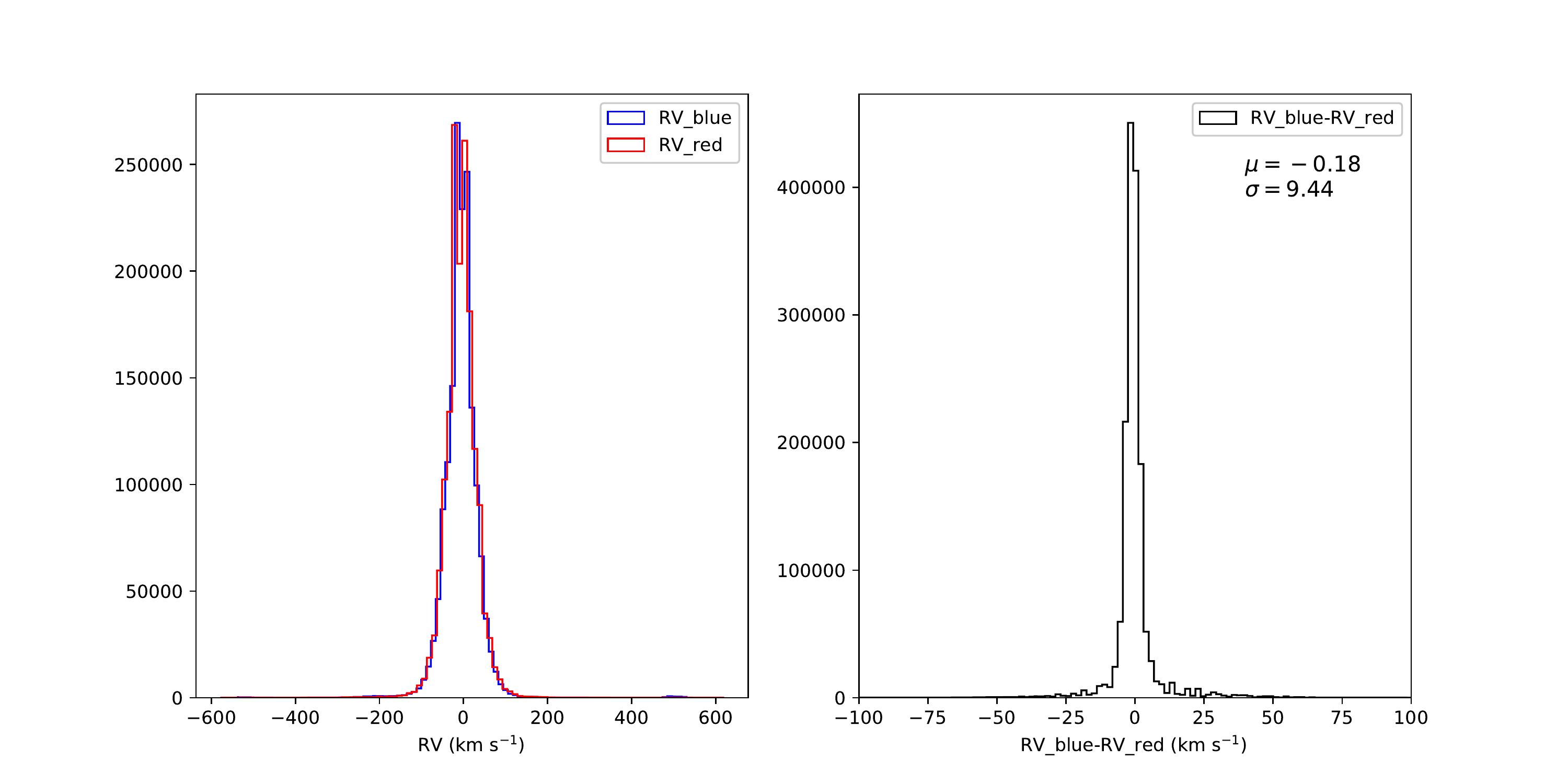}
\caption{The statistical difference of MRS RVs between blue and red waveband. The left panel shows the distribution of RVs derived from blue (in blue color) or red (in red color). The right panel is the difference between two RVs of blue and red.}
\label{Figure4}
\end{figure*}

It takes three steps to find the best fit template and corresponding Doppler shift for an observed MRS spectrum. The details are presented as follows:

\begin{enumerate}

\item  First, an observed spectrum is matched with all synthetic spectra of the grid that shifted with a course RV step of 40 $\text{km}~\text{s}^{-1}$  from -600 $\text{km}~\text{s}^{-1}$  to 600 $\text{km}~\text{s}^{-1}$. After this step, the rough RV of a spectrum has been decided.

\item  Second, the observed spectrum is matched with all synthetic spectra of the grid that shifted with a fine RV step of 1 $\text{km}~\text{s}^{-1}$ in $\pm{60}$ $\text{km}~\text{s}^{-1}$ around the optimal solution RV obtained in the above procedure. Then, a Gaussian fitting is done with ten points around the peak of the correlation function to determine the final RV estimation.

\item  For each spectrum, the blue and red wavebands are processed independently, and two RVs ($\text{RV}_{\text{blue}}$ and $\text{RV}_{\text{red}}$) are obtained.  Then the limit of S/N $>$10 is applied to both blue and red wavebands of all spectra, 1,594,956 out of 1,597,675 spectra (each has two single bands) are left. The additional cut is applied to remove the spectra with a large difference between $\text{RV}_{\text{blue}}$ and $\text{RV}_{\text{red}}$. Fig.\ref{Figure4} shows the difference of RVs between blue and red, and we can see that the 1$\sigma$ difference is 9.44$\text{km}~\text{s}^{-1}$. We use  $|\text{RV}_{\text{blue}}- \text{RV}_{\text{red}}| > 25~\text{km}~\text{s}^{-1}$  (close to 3 $\sigma$ difference) to carry out the cutting. Finally, 1,531,586 spectra (3,063,172 spectra if a single band is regarded as an individual spectrum) are left. 

\end{enumerate}

The first two steps are very computationally intensive for millions of spectra, we employ the distributed parallel computing platform Spark~\citep{Zaharia:2016:ASU:3013530.2934664} to deal with this challenge of  arduous computing task. The total time consuming of RV two-step calculation for LAMOST MRS 3,195,350 single-band spectra is about ten days using a computing cluster consisting of 15 PCs.

\section{Results}
\label{sect:results}

\subsection{Radial velocity zero point calibration}
The wavelength of LAMOSR MRS spectra is calibrated using Th-Ar and Sc arc lamps, which may shift with time from spanning a long time baseline. Time sequence study with multi-observations is one of the main goals of MRS, which requires the radial velocity zero point (RVZP) to be fixed using some RV standard stars (RV-STDs). To guarantee the stability of the RVZP, the RV-STDs which we plan to use should be proved that their RVs vary within a level of $100 \text{m}~\text{s}^{-1}$.  Many works provided RV-STDs from different survey projects, such as~\citet{2007sf2a.conf..459C,2010A&A...524A..10C} established a list of 1420 RV-STDs candidates, \citet{2013A&A...552A..64S,2018A&A...616A...7S} compiled a RV-STDs catalogue of 4813 stars for Gaia mission,  and \citet{2018AJ....156...90H} presented a catalogue of 18,080 RV-STDs selected from APOGEE data.

\begin{figure}[htbp]
\centering
\includegraphics[width=0.5\textwidth, angle=0]{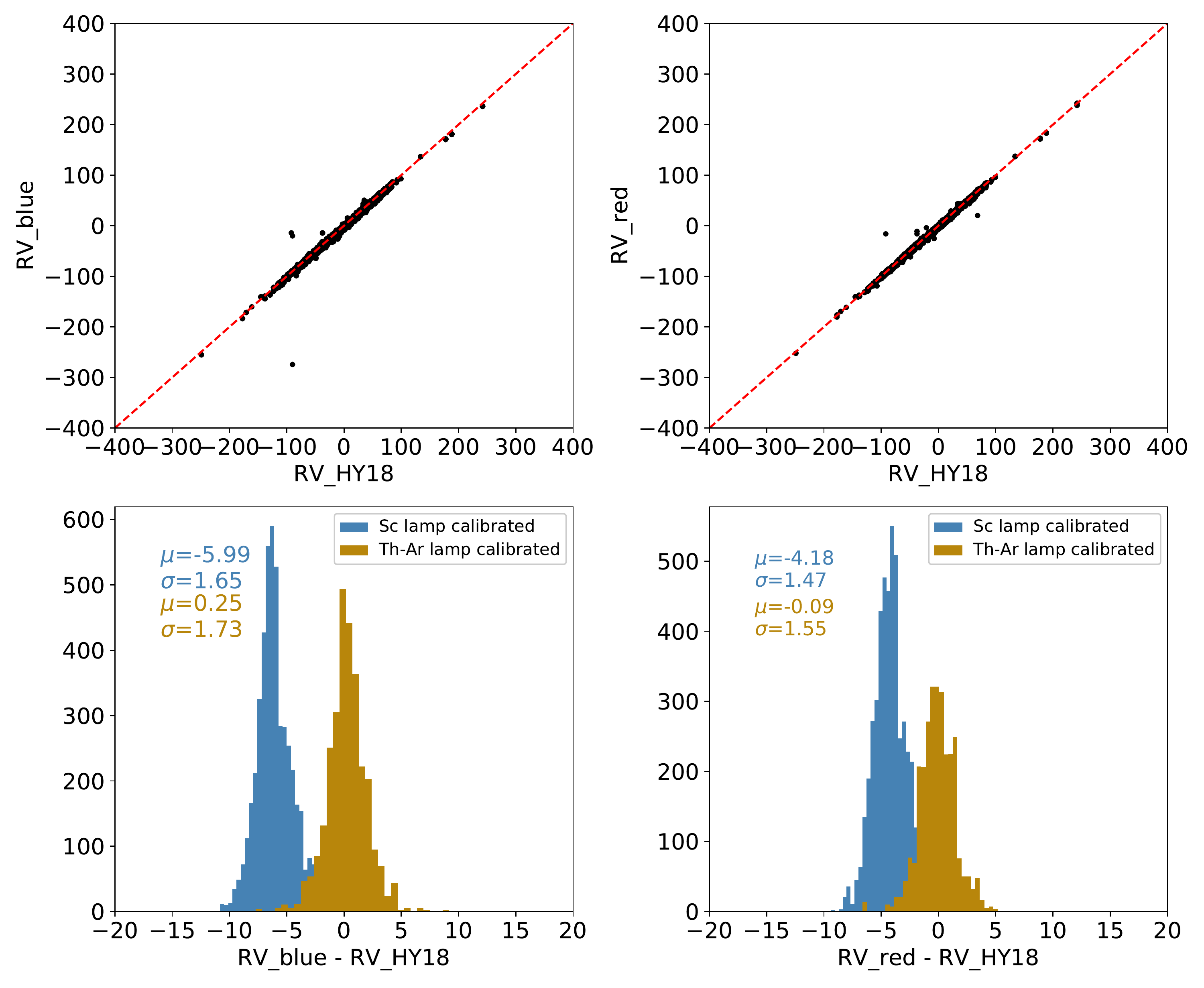}
\caption{Comparison between radial velocities ($\text{km}~\text{s}^{-1}$) of LAMOST MRS observations and RV-STDs from~\citet[hereafter, HY18]{2018AJ....156...90H}. For LAMOST blue waveband, the  comparison is shown in the left panels and for red waveband, it is shown in the right panels. Top panels show point-to-point comparisons, while bottom panels show histograms of their differences. During the MRS test observation, two different arc lamps, which are Sc lamp (blue) and Th-Ar lamp (brown),  were used to carry out the wavelength calibration for choosing a better lamp. }
\label{Figure5}
\end{figure}

We pick out 983 RV-STDs from \citet{2018AJ....156...90H} which have 7820 LAMOST MRS spectra covering all spectrographs and exposures. We compare the RV differences between the observed MRS spectra and corresponding RV-STDs.  Fig.\ref{Figure5} shows the overall RV offsets and dispersions for MRS RVs compared with RV-STDs. The RV zero (RVZP) point offset of the spectra with Sc lamp-based wavelength calibration is about -5.99 $\text{km}~\text{s}^{-1}$ for the blue waveband and -4.18 $\text{km}~\text{s}^{-1}$ for the red waveband. The RVZP offset based on Th-Ar lamp is, however, +0.25 $\text{km}~\text{s}^{-1}$ and -0.09 $\text{km}~\text{s}^{-1}$ for the blue and red waveband respectively. 

\begin{figure*}[htbp]
\centering
\includegraphics[width=\textwidth, angle=0]{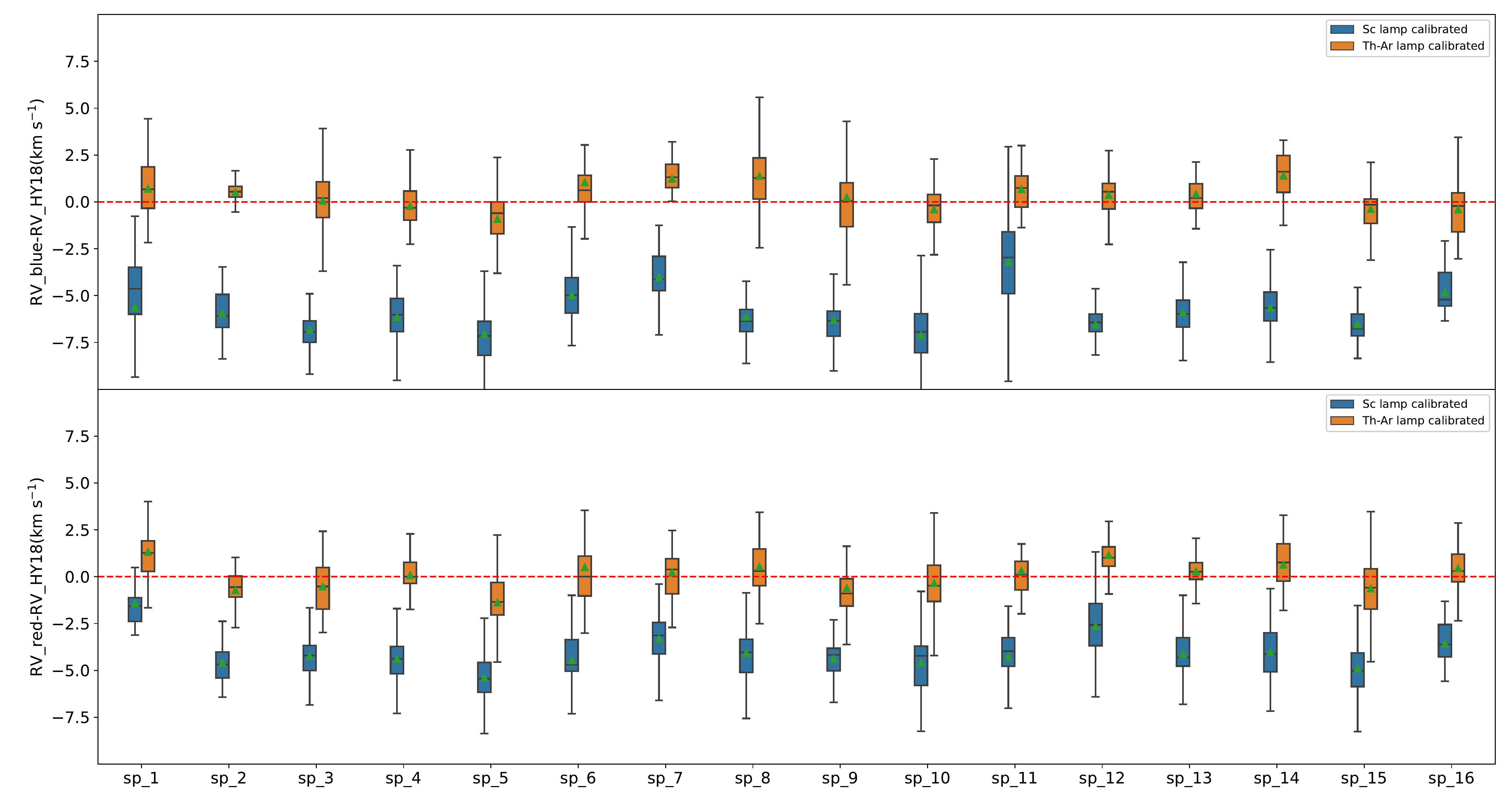}
\caption{Differences between radial velocities ($\text{km}~\text{s}^{-1}$) of LAMOST MRS observations and HY18 RV-STDs for individual spectrographs (spids), two wavelength coverages (blue in the top and red part in the bottom panel) and two kinds of arc lamps (brown for Th-Ar and blue for Sc lamp). The green triangles represent the means of the bars.}
\label{Figure6}
\end{figure*}

Considering the different situation for individual spectrographs, the RVZPs for each spectrograph with both Sc and Th-Ar lamp-based wavelength calibration is individually calculated and shown in the Fig.\ref{Figure6}. We can see that the RVZPs vary slightly with different spectrographs ("spid" in the figures indicates the ID of spectrograph). Before 19th Oct 2018, the Sc lamp is adapted to carry out the wavelength calibration for LAMOST MRS test observation spectra, while the Th-Ar lamp dominates later.  In fact, there is no essential difference between using these two lamps,  so the wavelength calibration of MRS spectra will be executed by  only using the Th-Ar lamp for future observations. 

\begin{figure}[htbp]
\centering
\includegraphics[width=0.5\textwidth, angle=0]{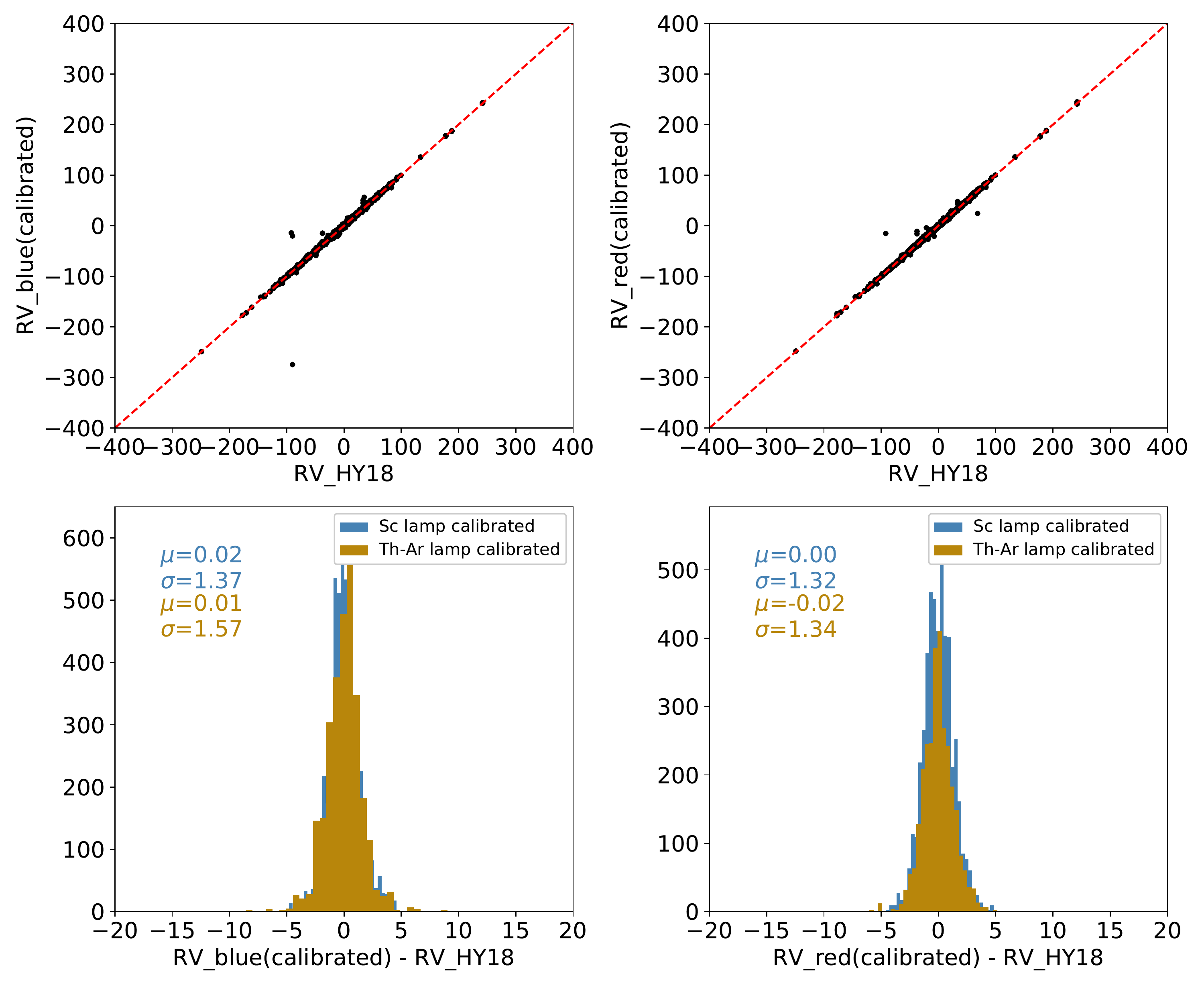}
\caption{Histograms of the differences between calibrated RVs of LAMOST MRS and RVs of HY18 RV-STDs for blue part spectra in the left panel and red part spectra in the right panel. The unit of RV is $\text{km}~\text{s}^{-1}$.}
\label{Figure7}
\end{figure}

Applying the RVZPs obtained from 7820 spectra of 983 RV-STDs to each exposure, spectrograph, and band, the RVs of all 1,594,956 two-band spectra are recalibrated to remove the offsets.  Fig.\ref{Figure7} shows the difference between recalibrated RVs of 7820 spectra and RVs from  HY18 RV-STDs. We plan to release both the RVs from two wavebands for all spectra before and after the RVZPs correction in the formal data release for different purpose of use. For example, RVs before the correction of RVZPs can be used to shift a spectrum to rest-frame while the calibrated RVs can be used in study kinematics. Here, we recommend the recalibrated RV of the blue part because of higher precision, and remind that using the RV of the red part should be cautious in case of strong H$\alpha$ emission in the absorption, referring to the ``$Flag_{red}=1$" in the catalog.

\subsection{Parameters of the best template}
\begin{figure}[htbp]
\centering
\includegraphics[width=0.5\textwidth, angle=0]{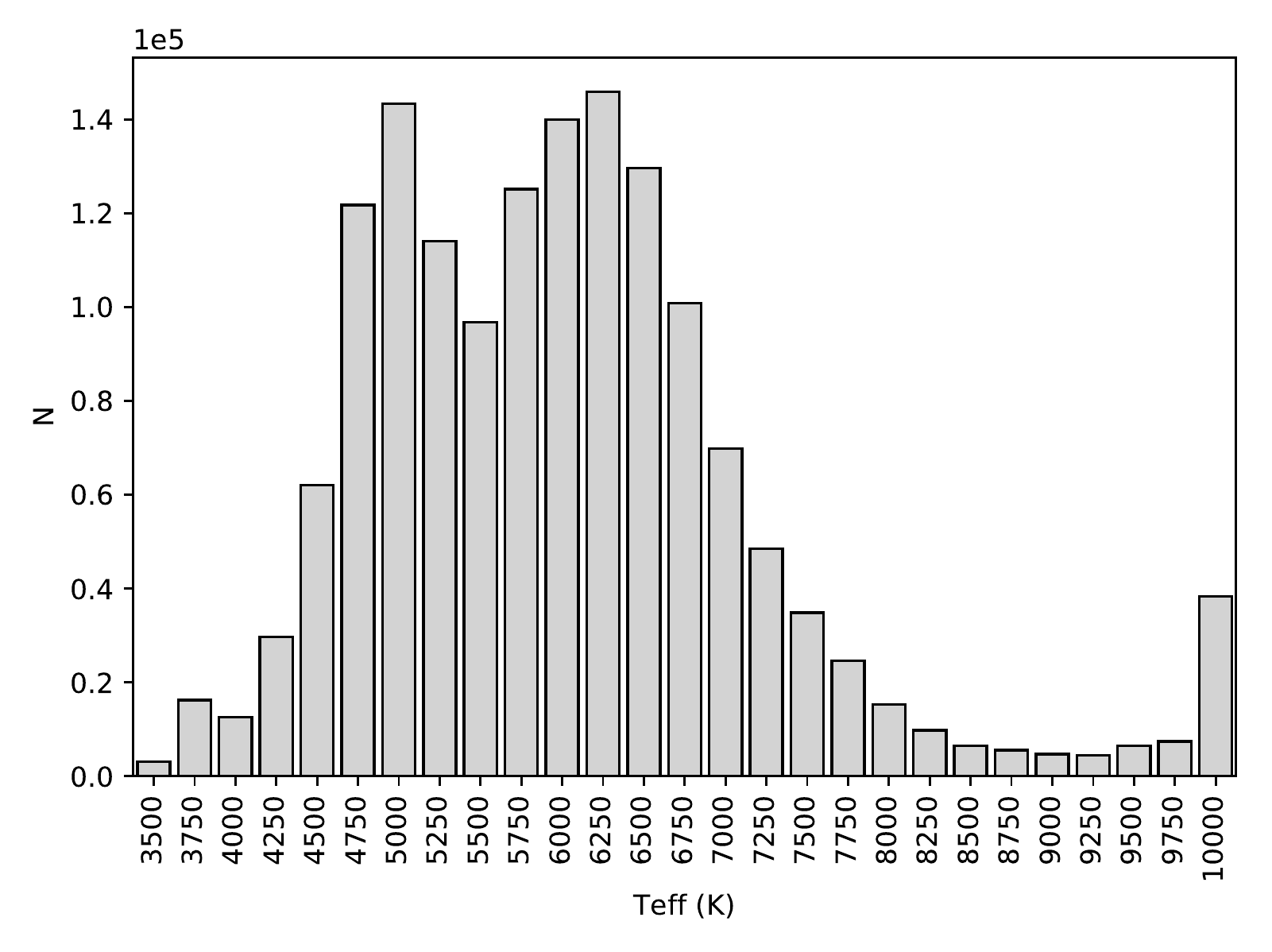}
\includegraphics[width=0.5\textwidth, angle=0]{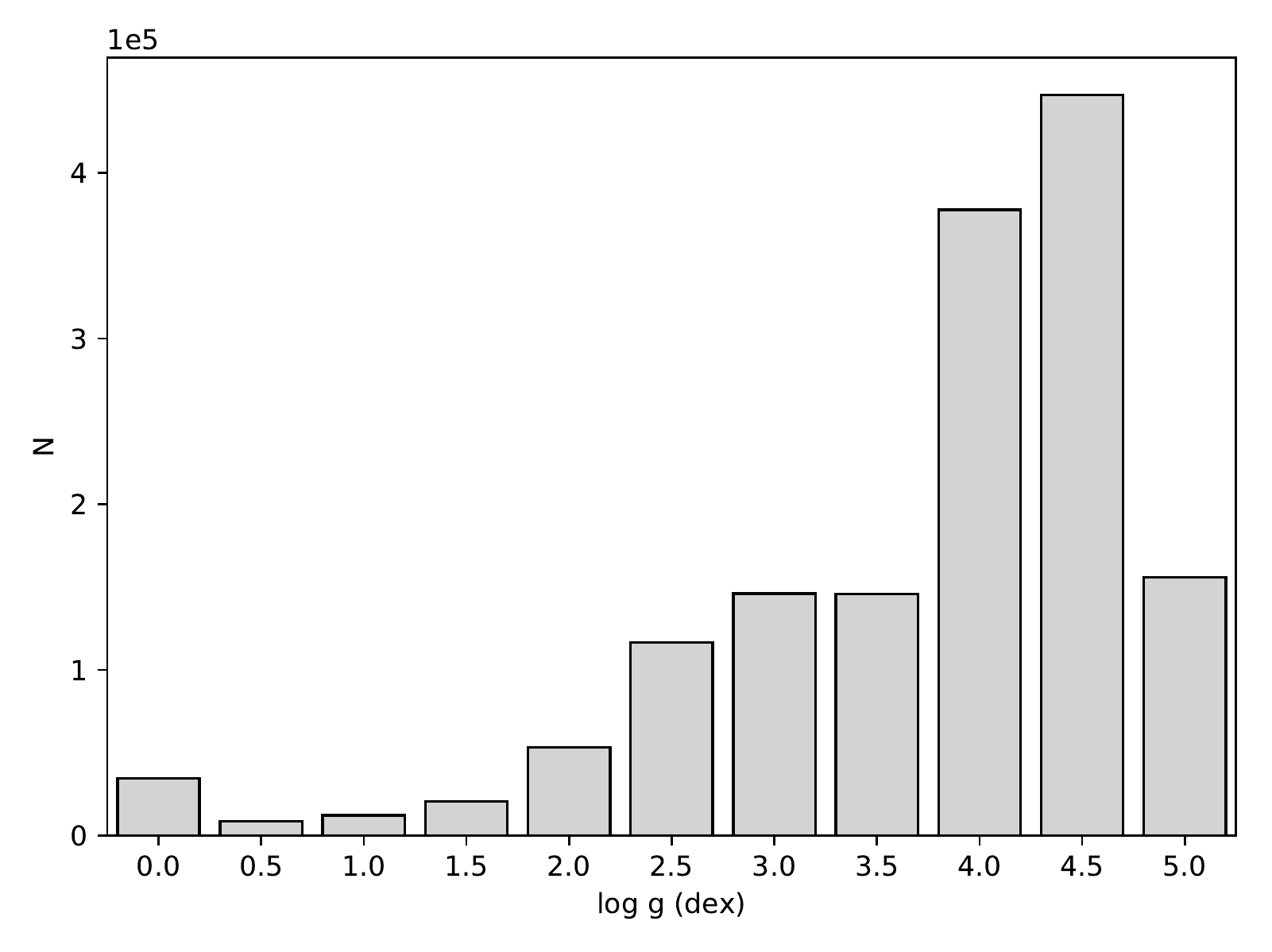}
\includegraphics[width=0.5\textwidth, angle=0]{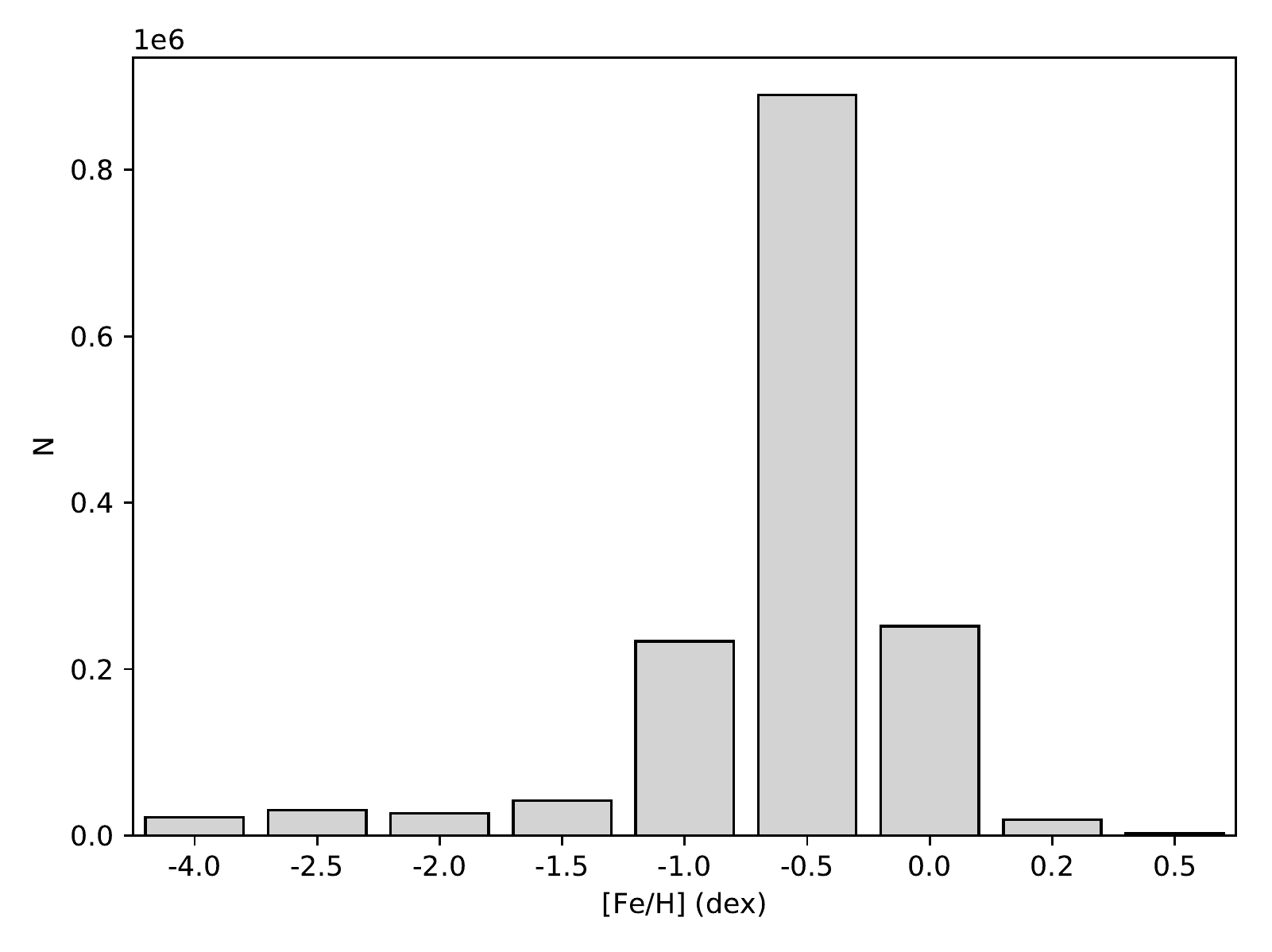}
\caption{Distribution of the effective temperatures (top panel), surface gravities (middle), and metallicities(bottom) of the templates}
\label{Figure8}
\end{figure}

To integrally understand the precision of RV measurement for LAMOST MRS, the spectral class of targets selected for the MRS test survey should have the same parameter span with the formal MRS survey. The stellar parameter distribution of the best matching templates used to determine RVs is shown in Figure~\ref{Figure8}. The effective temperature ranges from 3500K to 10000K  with two peaks at 5,000 K and 6,250 K. The surface gravities mainly concentrate around the value of 4.5 dex corresponding dwarfs, and a large part is also distributed between 2.5 to 3.5 dex corresponding giants and subgiants. The peak of metallicities ([Fe/H]) is -0.5 dex, and the number of stars fell sharply as [Fe/H] less than -1.0 dex and greater than 0 dex. This distribution of stellar parameters almost covers all kinds of stars for the future formal MRS survey. 

\subsection{Radial velocity accuracy}
To ensure the reliability and accuracy of the RVs of LAMOST MRS spectra, we employ some reference sets with RVs derived from medium/high-resolution observations for comparison and validation. To obtain reliable results, we only select LAMOST spectra with both $\text{S/N}_{\text{blue}} \geq 10$ and $\text{S/N}_{\text{red}} \geq 10$ for comparison. The reference sets we employed are as follows: 
\begin{enumerate}

\item The HY18 RV-STD catalogue~\citep{2018AJ....156...90H}. We cross-match the LAMOST MRS RV catalogue with HY18 RV-STD and get 1,106 common stars corresponding to 8,336 LAMOST spectra. 

\item The Gaia RV-STD catalogue~\citep{2018A&A...616A...7S}. We cross-match the LAMOST MRS RV catalogue with Gaia RV-STD and get 52 common stars corresponding to 326 LAMOST spectra. 

\item The Apache Point Observatory for Galactic Evolution Experiment DR14~\citep[APOGEE;][]{2015AJ....150..148H, 2017AJ....154...94M}. APOGEE is a spectroscopic survey with a median-high resolution of 22,500 in the near-infrared spectral range ($\lambda = 15700$ to $17500$ \AA). APOGEE Data Release 14~\citep[APOGEE DR14;][]{2018AJ....156..125H} has collected $\sim$277,000 spectra predominantly for giant stars. The complete APOGEE DR14 sample has 13,003 stars in common with the LAMOST MRS RV catalogue. We neglect the stars flagged by APOGEE as having large RV errors and finally a total of 89,741 spectra with repeated observations are left.

\item The Radial Velocity Experiment ~\citep[RAVE DR5;][]{2006AJ....132.1645S,2008AJ....136..421Z,2011AJ....141..187S,2013AJ....146..134K,2017AJ....153...75K}. RAVE is a spectroscopic survey with a same median resolution of R = 7500 as LAMOST medium resolution, covering the Ca\Rmnum{2} infra-red (IR) triplet region ($\lambda = 8410$ to $8795$ \AA) , and aim to measure radial velocities and stellar atmospheric parameters of one million stars using the 1.2 m UK Schmidt Telescope of the Anglo-Australian Observatory. RAVE Data Release 5~\citep[RAVE DR5;][]{2017AJ....153...75K} has presented 520,781 spectra of 457,588 unique stars. RAVE DR5 has 893 stars, corresponding to 6,449 spectra, in common with the LAMOST MRS RV catalogue after the exclusion of outliers. 

\item The Gaia Radial Velocity Spectrometer DR2 ~\citep[Gaia-RVS;][]{2004MNRAS.354.1223K,2018A&A...616A...5C}. Gaia-RVS on the European Space Agency’s Gaia mission is an integral-field spectrograph with resolving power of 11,500 covering the IR wavelength range within 8450-8720 \AA . Gaia Data Release 2~\citep[Gaia DR2;][]{2018A&A...616A...1G,2018A&A...616A...6S,2019A&A...622A.205K} published the first RVS measurements, which contains radial velocities for 7,224,631 stars. The Gaia RVS radial velocities catalogue has 152,734 stars in common with the LAMOST MRS RV catalogue corresponding to 868,663 spectra. 

\end{enumerate}

For each case, we use the common star subsets for comparison. The radial velocity residuals derived from the comparison of LAMOST MRS spectra with the reference sets are shown in Table.\ref{Tab:rv comparison samples}. LAMOST MRS shows a small offset of 0.028 $\text{km}~\text{s}^{-1}$ and 0.107 $\text{km}~\text{s}^{-1}$ with respect to two RV standard stars catalogues: HY18 RV-STD and Gaia RV-STD. For the comparison with the other three survey datasets (APOGEE DR14, RAVE and Gaia DR2), the offsets increase to 0.170 $\text{km}~\text{s}^{-1}$ $\sim$ 0.455 $\text{km}~\text{s}^{-1}$, which include the contributions from the error of both LAMOST MRS RVs and the other RV reference sets.

\begin{table}[!h]
\caption{Mean radial velocity residuals derived from the comparisons of LAMOST MRS data with the reference sets.}
\begin{center}
\label{Tab:rv comparison samples}
\begin{tabular}{lrrrrrr}
\hline\noalign{\smallskip}
\hline\noalign{\smallskip}
Catalogue &  $\mu_{\Delta\text{RV}}$ &  $\sigma_{\Delta\text{RV}}$  	&  $\text{N}_{\text{stars}}$  &  $\text{N}_{\text{spectra}}$\\
				& ($\text{km}~\text{s}^{-1}$) & ($\text{km}~\text{s}^{-1})$ & & &\\
\hline\noalign{\smallskip}
HY18 RV-STD		&    0.0277 		& 	1.4378		&    983			&  7,820					\\
Gaia RV-STD			&   -0.2797		&		1.6949		&  	46			&  261						\\
APOGEE	DR14		&   -0.2197 		&		2.7798		&  11,890		&  89,340					\\
RAVE DR5				&    0.0033 		&		4.1288		&  	834		&  5,934					\\
Gaia-RVS DR2		&   -0.3602 		&		2.1335		& 144,351	&  862,931				\\
\noalign{\smallskip}\hline
\end{tabular}
\end{center}
\end{table}

\begin{figure*}[htbp]
\centering
\includegraphics[width=\textwidth, height=0.33\textwidth, angle=0]{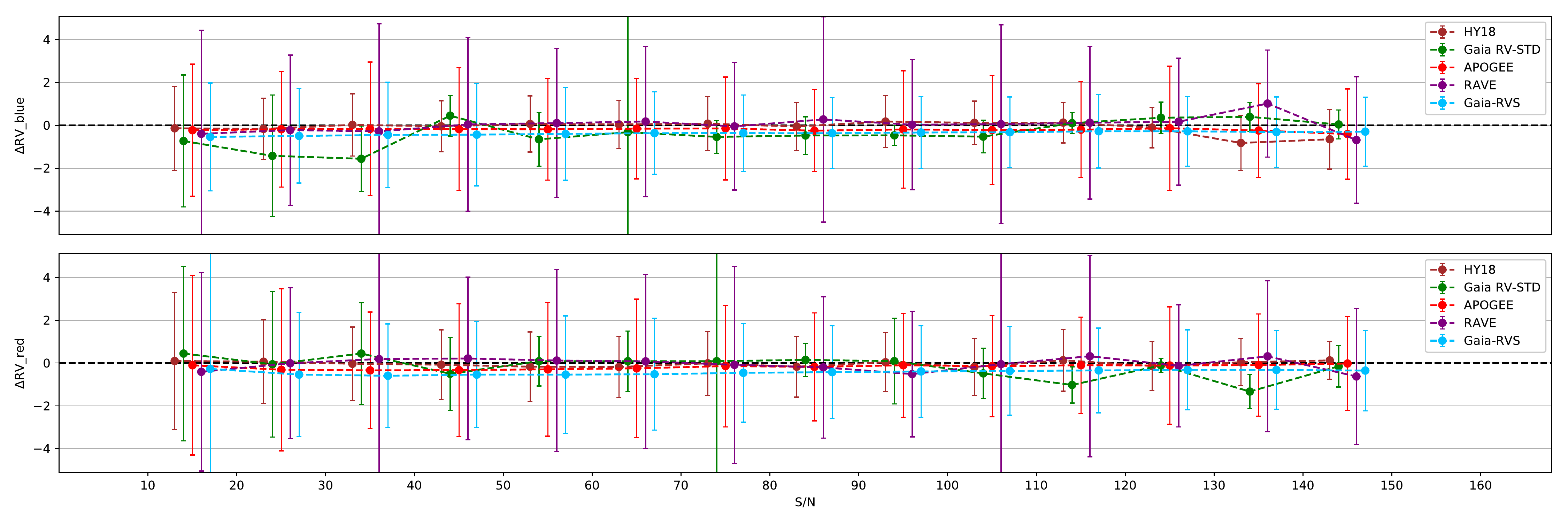}
\caption{Median radial velocity residuals of LAMOST MRS and reference sets as a function of the signal-to-noise S/N. Brown for comparison with HY18 RV-STD, green for Gaia RV-STD, red for APOGEE, purple for RAVE, and blue for Gaia-RVS. The top panel is for LAMOST RV derived from the spectra of blue part and the bottom is for the spectra of red part. Error-bars represent 1-$\sigma$ uncertainties. The step of S/N bins is 10.} 
\label{Figure9}
\end{figure*}

The accuracy estimated from the residuals of LAMOST RVs and reference RV data sets is shown as a function of signal-to-noise in Fig.\ref{Figure9}. LAMOST RVs shows neither significant offset nor trend with respect to the HY18 RV-STD or APOGEE with S/N increasing. It shows a nearly constant offset with respect to Gaia RV stars. For S/N higher than 100, the accuracy shows unexpected fluctuations because the number of spectra decreases rapidly. The same reason of little common star led to the residuals of RV of LAMOST and Gaia RV-STD is not stable as others. Both LAMOST MRS RV estimation and the reference sets contribute to the error bar in Fig.\ref{Figure9}. Most of the error of RV residuals decrease with S/N increasing. The RV differences between LAMOST and RAVE show larger than APOGEE and Gaia because the resolution of RAVE spectra is less than APOGEE's and Gaia-RVS's.   

\begin{figure*}[htbp]
\centering
\includegraphics[width=\textwidth, height=0.33\textwidth, angle=0]{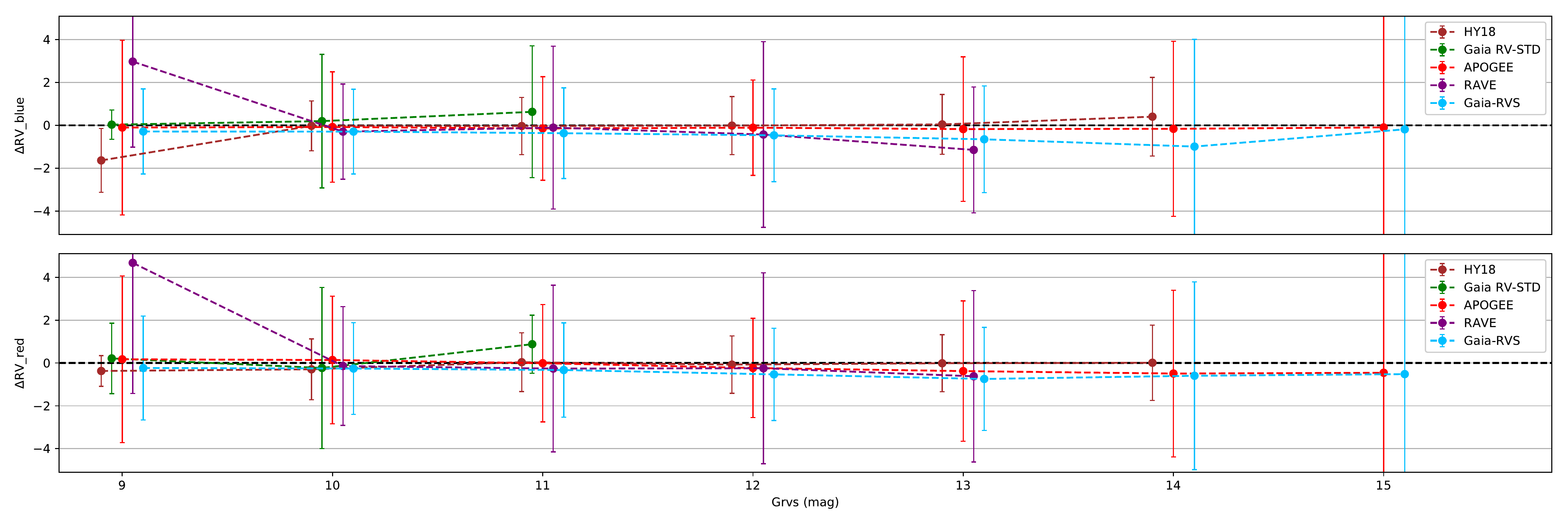}
\caption{Similar with Fig.\ref{Figure9}, median radial velocity residuals as a function of $\text{G}_{\text{RVS}}$ magnitude.  Error-bars represent 1-$\sigma$ uncertainties. The step of $\text{G}_{\text{RVS}}$ bins is 1 mag.}
\label{Figure10}
\end{figure*}

Fig.\ref{Figure10} shows the accuracy varies as a function of $\text{G}_{\text{RVS}}$. The RV residuals stay at a very low level for stars with $\text{G}_{\text{RVS}}$ of 9 even to 15 mag, except for that the RV residuals between LAMOST and RAVE become upturned at $\text{G}_{\text{RVS}}$ bright end. According to the other reference sets, the reason for "upturning" probably comes from the inaccuracy of RAVE RV estimation. Besides, LAMOST radial velocities  exhibit an inconsistent with Gaia RV-STD when $\text{G}_{\text{RVS}}$ at 11 mag, the median RV residuals reaching about 0.9 $\text{km}~\text{s}^{-1}$ at $\text{G}_{\text{RVS}} $ = 11 mag for blue part.

\begin{figure*}[htbp]
\centering
\includegraphics[width=\textwidth, height=0.33\textwidth, angle=0]{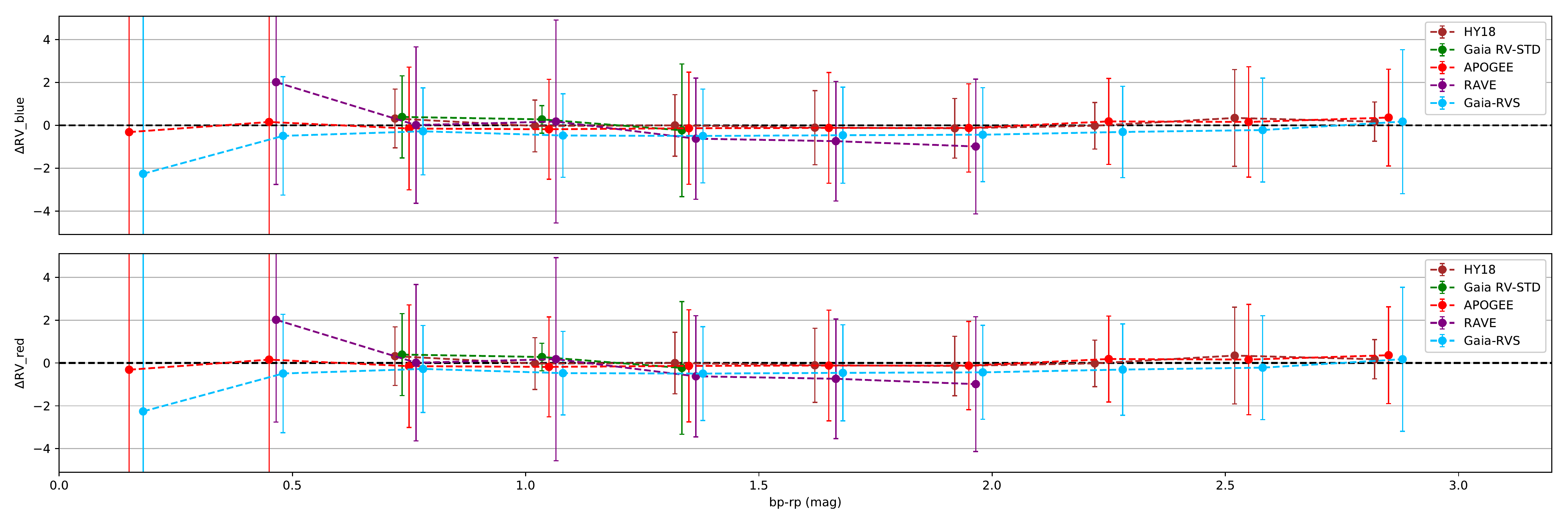}
\caption{Similar with Fig.\ref{Figure9}, median radial velocity residuals as a function of color index \textit{bp-rp}.  Error-bars represent 1-$\sigma$ uncertainties. The step of \textit{bp-rp} bins is 0.3 mag.}
\label{Figure11}
\end{figure*}

Fig.\ref{Figure11} shows the accuracy varies as a function of the color index \textit{bp-rp} employed from Gaia DR2 photometric catalogue. LAMOST MRS RVs are well consistent with the reference catalogues for the stars with \textit{bp-rp} higher than 0.5 mag. It needs to point out that RVAE RVs are inconsistent with LAMOST RVs and other RV reference sets in many situations, such as stars with \textit{bp-rp} around 0.5 and 2.0 mag.

\subsection{Radial velocity precision}

\begin{figure*}[htbp]
\centering
\includegraphics[width=\textwidth, height=0.33\textwidth, angle=0]{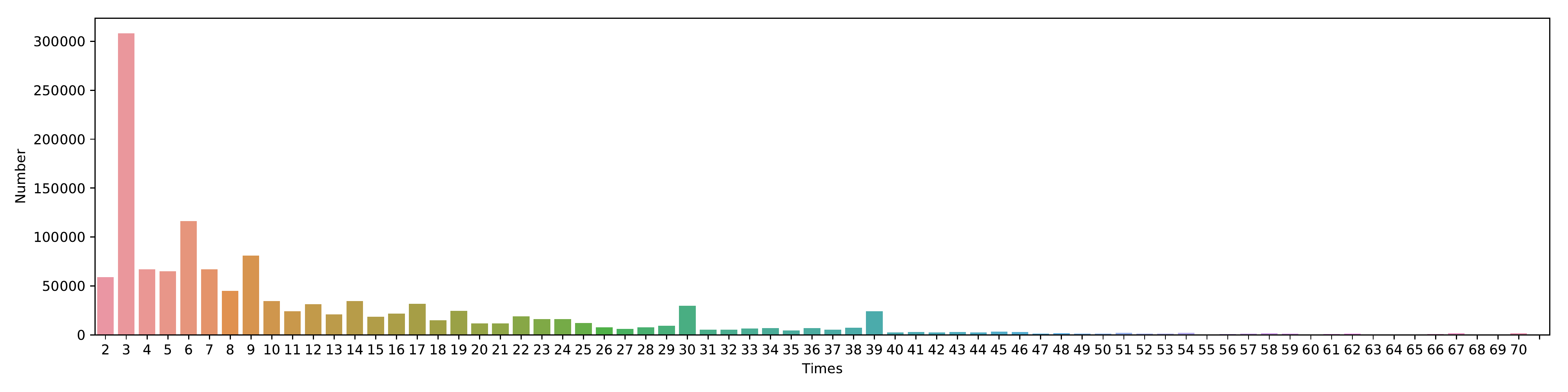}
\caption{Distribution of the numbers of repeated LAMOST MRS observations with measurable radial velocities.}
\label{Figure12}
\end{figure*}

The precision of RVs is estimated based on the RV measurements of multiple observations in different epoch for the same stars. Fig.\ref{Figure12} shows the distribution of the numbers of repeated observations with measurable radial velocities. The statistical estimator used to access the precision is: 
\\
\begin{equation}
\epsilon = \sqrt{\frac{\text{N}}{\text{N-1}} \sum\limits_{i=0}^{\text{N}}(\text{RV}_{i}-\overline{\text{RV}})^{2}},
\end{equation}
\\
where $\text{N}$ is the number of repeated observations and $\overline{\text{RV}}$ is the mean value of $\text{RV}_{i}$.

\begin{figure*}[htbp]
\centering
\includegraphics[width=\textwidth, angle=0]{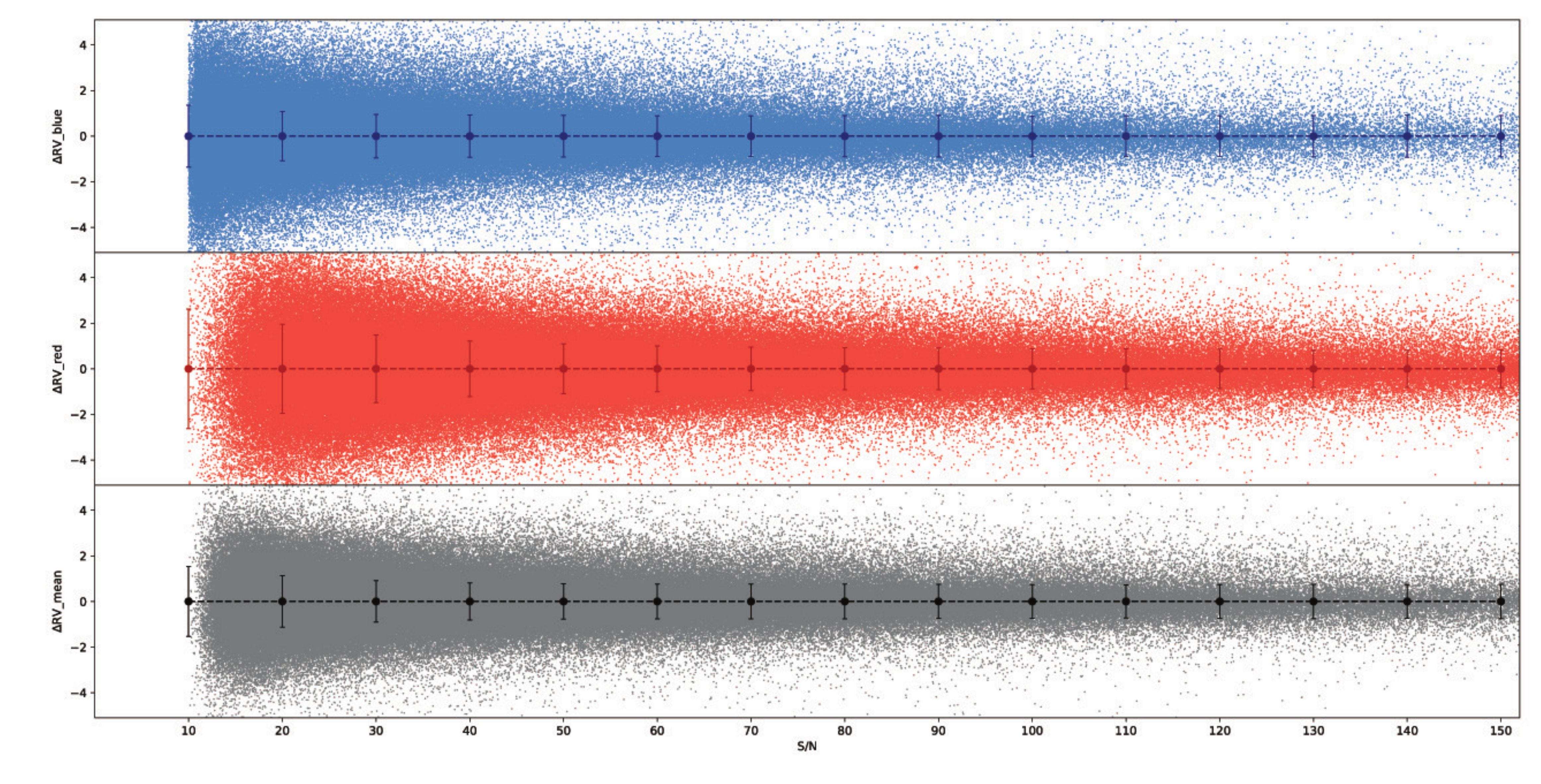}
\caption{The precision of RV derived from repeated observations as a function of S/N. The top panel shows the precision for RV derived from blue part $\text{RV}_{\text{blue}}$, the middle panel shows the precision from the red part ($\text{RV}_{\text{red}}$) and the bottom shows the mean value of $\text{RV}_{\text{blue}}$ and $\text{RV}_{\text{red}}$. The error-bar step of S/N is 10.}
\label{Figure13}
\end{figure*}

The main factor affecting the precision of RV measurement of a spectrum is its signal-to-noise ($\text{S/N}$). Fig.\ref{Figure13} shows the precision of radial velocities as a function of $\text{S/N}$  for the LAMOST MRS spectra with $\text{S/N} \geq 10$. In  Fig.\ref{Figure13}, the top panel displays RVs measured by the blue parts of the spectra ($\text{RV}_{\text{blue}}$), the middle panel is the $\text{RV}_{\text{red}}$ derived from the red part and the bottom is the mean values of $\text{RV}_{\text{blue}}$ and $\text{RV}_{\text{red}}$. The precision improves as the signal-to-noise increases. For spectra of blue parts with $\text{S/N}_{\text{blue}}$ equal to 10, the precision is 1.36 $\text{km}~\text{s}^{-1}$ and improves to less than 1.0 $\text{km}~\text{s}^{-1}$ when $\text{S/N}_{\text{blue}}$ increases to higher than 30. The precision of $\text{RV}_{\text{blue}}$ is better than $\text{RV}_{\text{red}}$, not only because there are more absorption lines in the blue part of the most type of spectra than that in the red part, but also the lines in blue part are sharper. 

\subsection{a RV catalog for LAMOST MRS test survey}

We provide a RV catalog for the LAMOST MRS test survey containing 1,594,956 spectra. The information in the catalogue includes: Gaia  identier
(Gaia source id), the identifier for correspond-ing star (starid),  LAMOST spectral identifier \emph{medid}, observation information ($\text{obsdate}$, $\text{spid}$, $\text{lamp}$), right ascension (RA), declination (Dec), signal-to-noise of the spectra (S/N), radial velocities without RVZP correction($\text{RV}_{\text{blue}}$, $\text{RV}_{\text{red}}$), corresponding RV errors ($\text{ERV}_{\text{blue}}$, $\text{ERV}_{\text{red}}$), radial velocities with RVZP corrected  ($\text{RV}_{\text{blue,cali}}$, $\text{RV}_{\text{red,cali}}$), correlation coefficient between the observation and the best-fit template (\emph{corr}), effective temperature ($\textit{T}_{eff}$), surface gravity (log $g$) and metallicities ([Fe/H]) of the best-fit template, the quality flag for accessing radial velocities, the corresponding Gaia photometric magnitude ($\text{G}_{\text{RVS}}$), color index (\textit{bp-rp}). A description of columns of the LAMOST RV catalogue is shown in Tab~\ref{Tab:description}. The full catalogue can be accessed on-line at http://paperdata.china-vo.org/LAMOST/ LAMOST MRS RV.csv.

\begin{table*}[!h]
\caption{Description of the columns of the LAMOST MRS RV catalogue.}
\begin{center}
\label{Tab:description}
\begin{tabular}{lll}
\hline\noalign{\smallskip}
\hline\noalign{\smallskip}
Col. &  Name  	&  Description \\
\hline\noalign{\smallskip}
1  & Gaia$\_{\text{source}\_\text{id}}$ & Gaia source id by cross-matching Gaia DR2.\\
2 &\text{starid} & id for corresponding star based on the RA and Dec, with the form of LAMOST Jdddmmss ddmmss.\\
3 & $\text{medid}$ & LAMOST spectral ID, with the form of Date-PlateID-SpectrographID-FiberID-MJM-PiplineVersion. \\
4 &$\text{medid}_{\text{blue}}$ &  LAMOST spectral ID for blue part. \\
5 &$\text{medid}_{\text{red}}$ &  LAMOST spectral ID for red part. \\
6 & $\text{obsdate}$ &  Date of the observation. \\
7 & $\text{spid}$ & Spectrograph ID. \\
8 & $\text{lamp}$ &  Lamp used for wavelength calibration. \\
9 & $\text{RA}$ &  Right ascension of J2000. ($^{\circ}$) \\
10 & $\text{Dec}$& Declination of J2000. ($^{\circ}$) \\
11 &$\text{S/N}_{\text{blue}}$ & Signal-to-noise of blue part. \\
12 &$\text{S/N}_{\text{red}}$ & Signal-to-noise of red part. \\
13 &$\text{RV}_{\text{blue}}$ & RV measured from blue part without zero point corrected. ($\text{km}~\text{s}^{-1}$) \\
14 &$\text{ERV}_{\text{blue}}$ & Error of RV measured from blue part. ($\text{km}~\text{s}^{-1}$) \\
15 &$\text{RV}_{\text{red}}$ & RV measured from red part without zero point corrected. ($\text{km}~\text{s}^{-1}$) \\
16 &$\text{ERV}_{\text{red}}$ & Error of RV measured from red part. ($\text{km}~\text{s}^{-1}$) \\
17 &$\text{RV}_{\text{blue,cali}}$& RV with zero point corrected for blue part. ($\text{km}~\text{s}^{-1}$) \\
18 &$\text{RV}_{\text{red,cali}}$ & RV with zero point corrected for red part. ($\text{km}~\text{s}^{-1}$) \\
19 &$\emph{corr}_{\text{blue}}$ & Correlation coefficient between blue part and the best-fit template. \\
20 &$\emph{corr}_{\text{red}}$ & Correlation coefficient between red part and the best-fit template. \\
21 &$\textit{T}_{\text{eff}}\_{\text{blue}}$ & Effective temperature of the best-fit template of blue part. (K) \\
22 &log $g\_{\text{blue}}$ & Surface gravity of the best-fit template of blue part. (dex) \\
23 &[Fe/H]$\_{\text{blue}}$ & [Fe/H] of the best-fit template of blue part. (dex) \\
24 &$\textit{T}_{\text{eff}}\_{\text{red}}$ & Effective temperature of the best-fit template of red part. (K) \\
25 &log $g\_{\text{red}}$ & Surface gravity of the best-fit template of red part. (dex) \\
26 &[Fe/H]$\_{\text{red}}$ & [Fe/H] of the best-fit template of red part. (dex) \\
27 &Flag$\_{\text{blue}}$ &  Quality flag for accessing radial velocity derived from blue part. 0 for good, else 1 for bad.  \\
28 &Flag$\_{\text{red}}$ & Quality flag for accessing radial velocity derived from blue part. 0 for good, else 1 for bad.  \\
29 &$\text{G}_{\text{RVS}}$ & G magnitude employed from Gaia DR2 photometric catalogue. (mag) \\
30 &$\textit{bp-rp}$ & bp-rp color index employed from Gaia DR2 photometric catalogue. (mag) \\

\noalign{\smallskip}\hline
\end{tabular}
\end{center}
\tablecomments{The full catalogue can be accessed on-line at http://paperdata.china-vo.org/LAMOST/ LAMOST MRS RV.csv.}
\end{table*}

\section{Summary }
\label{sect:summary}

The test observation for the LAMOST medium-resolution (R=7500) survey (MRS) began on 1st Sep 2017, and a total of 1,594,956 spectra (each has two bands) of stars with S/N higher than 10 has been collected till 31st Dec 2017. We measure radial velocities (RVs) for the dataset by cross-correlation method, and corrected the zero point of the RVs based on a set of standard stars (RV-STD). Then, we evaluate the properties of RVs of LAMOST spectra by comparing with five reference sets: HY18 RV-STD, Gaia RV-STD, APOGEE DR14, RAVE DR5, and GAIA DR2. The RVs of the LAMOST MRS test survey perform well,  the accuracy is 0.03 $\text{km}~\text{s}^{-1}$ comparing to HY18 RV-STD and 0.28 $\text{km}~\text{s}^{-1}$ comparing to Gaia RV-STD. The precision of the measurement can be obtained through repeat observation, and mainly dominated by signal-to-noise, for example, the RV precision is from 1.36 $\text{km}~\text{s}^{-1}$ to 0.91 $\text{km}~\text{s}^{-1}$ as the signal-to-noise levels from 10 to 50. We provide a RV catalogue which is available on the website \url{http://paperdata.china-vo.org/LAMOST/LAMOST_MRS_RV.csv}. In the future formal MRS, we will put enough RV-STDs in the input catalog of LAMOST MRS observations to correct the zero point for each spectrometer in each exposure.

\section*{Acknowledgement}
This work is supported by the National Key Basic Research Program of China (Grant No. 2014CB845700), the National Natural Science Foundation of China (Grant No. 11390371), China Scholarship Council, Key Research Program of Frontier Sciences, CAS (Grant No. QYZDY-SSW-SLH007). The Guo Shou Jing Telescope (the Large Sky Area Multi-Object Fiber Spectroscopic Telescope, LAMOST) is a National Major Scientific Project built by the Chinese Academy of Sciences. Funding for the project has been provided by the National Development and Reform Commission. LAMOST is operated and managed by National Astronomical Observatories, Chinese Academy of Sciences.

\software{
		Numpy\citep{oliphant_guide_2006},
		Scipy\citep{jones2001scipy},
		Matplotlib\citep{Hunter:2007},
		Pandas\citep{pythonpandas},
		Astropy\citep{2013A&A...558A..33A},
		Spark\citep{Zaharia:2016:ASU:3013530.2934664}
		}
\bibliography{reference}

\begin{thebibliography}{}
\expandafter\ifx\csname natexlab\endcsname\relax\def\natexlab#1{#1}\fi
\providecommand{\url}[1]{\href{#1}{#1}}
\providecommand{\dodoi}[1]{doi:~\href{http://doi.org/#1}{\nolinkurl{#1}}}
\providecommand{\doeprint}[1]{\href{http://ascl.net/#1}{\nolinkurl{http://ascl.net/#1}}}
\providecommand{\doarXiv}[1]{\href{https://arxiv.org/abs/#1}{\nolinkurl{https://arxiv.org/abs/#1}}}

\bibitem[{{Astropy Collaboration} {et~al.}(2013){Astropy Collaboration},
  {Robitaille}, {Tollerud}, {Greenfield}, {Droettboom}, {Bray}, {Aldcroft},
  {Davis}, {Ginsburg}, {Price-Whelan}, {Kerzendorf}, {Conley}, {Crighton},
  {Barbary}, {Muna}, {Ferguson}, {Grollier}, {Parikh}, {Nair}, {Unther},
  {Deil}, {Woillez}, {Conseil}, {Kramer}, {Turner}, {Singer}, {Fox}, {Weaver},
  {Zabalza}, {Edwards}, {Azalee Bostroem}, {Burke}, {Casey}, {Crawford},
  {Dencheva}, {Ely}, {Jenness}, {Labrie}, {Lim}, {Pierfederici}, {Pontzen},
  {Ptak}, {Refsdal}, {Servillat}, \& {Streicher}}]{2013A&A...558A..33A}
{Astropy Collaboration}, {Robitaille}, T.~P., {Tollerud}, E.~J., {et~al.} 2013,
  \aap, 558, A33, \dodoi{10.1051/0004-6361/201322068}

\bibitem[{{Bouchy} {et~al.}(2001){Bouchy}, {Pepe}, \&
  {Queloz}}]{2001A&A...374..733B}
{Bouchy}, F., {Pepe}, F., \& {Queloz}, D. 2001, \aap, 374, 733,
  \dodoi{10.1051/0004-6361:20010730}

\bibitem[{{Britavskiy} {et~al.}(2018){Britavskiy}, {Pancino}, {Tsymbal},
  {Romano}, \& {Fossati}}]{2018MNRAS.474.3344B}
{Britavskiy}, N., {Pancino}, E., {Tsymbal}, V., {Romano}, D., \& {Fossati}, L.
  2018, \mnras, 474, 3344, \dodoi{10.1093/mnras/stx2944}

\bibitem[{{Castelli} \& {Kurucz}(2004)}]{2004astro.ph..5087C}
{Castelli}, F., \& {Kurucz}, R.~L. 2004, ArXiv Astrophysics e-prints

\bibitem[{{Crifo} {et~al.}(2010){Crifo}, {Jasniewicz}, {Soubiran}, {Katz},
  {Siebert}, {Veltz}, \& {Udry}}]{2010A&A...524A..10C}
{Crifo}, F., {Jasniewicz}, G., {Soubiran}, C., {et~al.} 2010, \aap, 524, A10,
  \dodoi{10.1051/0004-6361/201015315}

\bibitem[{{Crifo} {et~al.}(2007){Crifo}, {Jasniewicz}, {Soubiran},
  {Hestroffer}, {Siebert}, {Guerrier}, {Katz}, {Th{\'e}venin}, \&
  {Turon}}]{2007sf2a.conf..459C}
{Crifo}, F., {Jasniewicz}, G., {Soubiran}, C., {et~al.} 2007, in SF2A-2007:
  Proceedings of the Annual meeting of the French Society of Astronomy and
  Astrophysics, ed. J.~{Bouvier}, A.~{Chalabaev}, \& C.~{Charbonnel}, 459

\bibitem[{{Cropper} {et~al.}(2018){Cropper}, {Katz}, {Sartoretti}, {Prusti}, \&
  others.}]{2018A&A...616A...5C}
{Cropper}, M., {Katz}, D., {Sartoretti}, P., {Prusti}, T., \& others. 2018,
  \aap, 616, A5, \dodoi{10.1051/0004-6361/201832763}

\bibitem[{{Cui} {et~al.}(2012){Cui}, {Zhao}, {Chu}, {Li}, \&
  others.}]{2012RAA....12.1197C}
{Cui}, X.-Q., {Zhao}, Y.-H., {Chu}, Y.-Q., {Li}, G.-P., \& others. 2012,
  Research in Astronomy and Astrophysics, 12, 1197,
  \dodoi{10.1088/1674-4527/12/9/003}

\bibitem[{{De Silva} {et~al.}(2015){De Silva}, {Freeman}, {Bland-Hawthorn}, \&
  others.}]{2015MNRAS.449.2604D}
{De Silva}, G.~M., {Freeman}, K.~C., {Bland-Hawthorn}, J., \& others. 2015,
  MNRAS, 449, 2604, \dodoi{10.1093/mnras/stv327}

\bibitem[{{Deng} {et~al.}(2012){Deng}, {Newberg}, {Liu}, {Carlin}, {Beers},
  {Chen}, {Chen}, {Christlieb}, {Grillmair}, {Guhathakurta}, {Han}, {Hou},
  {Lee}, {L{\'e}pine}, {Li}, {Liu}, {Pan}, {Sellwood}, {Wang}, {Wang}, {Yang},
  {Yanny}, {Zhang}, {Zhang}, {Zheng}, \& {Zhu}}]{2012RAA....12..735D}
{Deng}, L.-C., {Newberg}, H.~J., {Liu}, C., {et~al.} 2012, Research in
  Astronomy and Astrophysics, 12, 735, \dodoi{10.1088/1674-4527/12/7/003}

\bibitem[{{Gaia Collaboration} {et~al.}(2018){Gaia Collaboration}, {Brown},
  {Vallenari}, {Prusti}, {de Bruijne}, {Babusiaux}, {Bailer-Jones}, {Biermann},
  {Evans}, {Eyer}, \& et~al.}]{2018A&A...616A...1G}
{Gaia Collaboration}, {Brown}, A.~G.~A., {Vallenari}, A., {et~al.} 2018, \aap,
  616, A1, \dodoi{10.1051/0004-6361/201833051}

\bibitem[{{Geller} {et~al.}(2015){Geller}, {Latham}, \&
  {Mathieu}}]{2015AJ....150...97G}
{Geller}, A.~M., {Latham}, D.~W., \& {Mathieu}, R.~D. 2015, \aj, 150, 97,
  \dodoi{10.1088/0004-6256/150/3/97}

\bibitem[{{Gilmore} {et~al.}(2012){Gilmore}, {Randich}, {Asplund}, {Binney},
  {Bonifacio}, {Drew}, {Feltzing}, {Ferguson}, {Jeffries}, {Micela}, \&
  et~al.}]{2012Msngr.147...25G}
{Gilmore}, G., {Randich}, S., {Asplund}, M., {et~al.} 2012, The Messenger, 147,
  25

\bibitem[{{He{\l}miniak} {et~al.}(2018){He{\l}miniak}, {Tokovinin},
  {Niemczura}, {Paw{\l}aszek}, {Yanagisawa}, {Brahm}, {Espinoza}, {Ukita},
  {Ratajczak}, {Hempel}, {Jord{\'a}n}, {Konacki}, {Sybilski}, {Koz{\l}owski},
  {Litwickim}, \& {Tamura}}]{2018arXiv181204319H}
{He{\l}miniak}, K.~G., {Tokovinin}, A., {Niemczura}, E., {et~al.} 2018, arXiv
  e-prints.
\newblock \doarXiv{1812.04319}

\bibitem[{{Holtzman} {et~al.}(2015){Holtzman}, {Shetrone}, {Johnson}, {Allende
  Prieto}, {Anders}, {Andrews}, {Beers}, {Bizyaev}, {Blanton}, {Bovy},
  {Carrera}, {Chojnowski}, {Cunha}, {Eisenstein}, {Feuillet}, {Frinchaboy},
  {Galbraith-Frew}, {Garc{\'{\i}}a P{\'e}rez}, {Garc{\'{\i}}a-Hern{\'a}ndez},
  {Hasselquist}, {Hayden}, {Hearty}, {Ivans}, {Majewski}, {Martell},
  {Meszaros}, {Muna}, {Nidever}, {Nguyen}, {O'Connell}, {Pan}, {Pinsonneault},
  {Robin}, {Schiavon}, {Shane}, {Sobeck}, {Smith}, {Troup}, {Weinberg},
  {Wilson}, {Wood-Vasey}, {Zamora}, \& {Zasowski}}]{2015AJ....150..148H}
{Holtzman}, J.~A., {Shetrone}, M., {Johnson}, J.~A., {et~al.} 2015, AJ, 150,
  148, \dodoi{10.1088/0004-6256/150/5/148}

\bibitem[{{Holtzman} {et~al.}(2018){Holtzman}, {Hasselquist}, {Shetrone},
  {Cunha}, {Allende Prieto}, {Anguiano}, {Bizyaev}, {Bovy}, {Casey},
  {Edvardsson}, {Johnson}, {J{\"o}nsson}, {Meszaros}, {Smith}, {Sobeck},
  {Zamora}, {Chojnowski}, {Fernandez-Trincado}, {Garcia-Hernandez}, {Majewski},
  {Pinsonneault}, {Souto}, {Stringfellow}, {Tayar}, {Troup}, \&
  {Zasowski}}]{2018AJ....156..125H}
{Holtzman}, J.~A., {Hasselquist}, S., {Shetrone}, M., {et~al.} 2018, AJ, 156,
  125, \dodoi{10.3847/1538-3881/aad4f9}

\bibitem[{{Huang} {et~al.}(2018){Huang}, {Liu}, {Chen}, {Zhang}, {Yuan},
  {Xiang}, {Wang}, \& {Tian}}]{2018AJ....156...90H}
{Huang}, Y., {Liu}, X.-W., {Chen}, B.-Q., {et~al.} 2018, \aj, 156, 90,
  \dodoi{10.3847/1538-3881/aacda5}

\bibitem[{Hunter(2007)}]{Hunter:2007}
Hunter, J. 2007, Computing In Science \& Engineering, 9, 90,
  \dodoi{10.1109/MCSE.2007.55}

\bibitem[{Jones {et~al.}(2001--)Jones, Oliphant, Peterson,
  {et~al.}}]{jones2001scipy}
Jones, E., Oliphant, T., Peterson, P., {et~al.} 2001--, {SciPy}: Open source
  scientific tools for {Python}.
\newblock \url{http://www.scipy.org/}

\bibitem[{{Katz} {et~al.}(2004){Katz}, {Munari}, {Cropper}, {Zwitter}, \&
  others.}]{2004MNRAS.354.1223K}
{Katz}, D., {Munari}, U., {Cropper}, M., {Zwitter}, T., \& others. 2004, MNRAS,
  354, 1223, \dodoi{10.1111/j.1365-2966.2004.08282.x}

\bibitem[{{Katz} {et~al.}(2019){Katz}, {Sartoretti}, {Cropper}, {Panuzzo},
  {Seabroke}, {Viala}, {Benson}, {Blomme}, {Jasniewicz}, {Jean-Antoine},
  {Huckle}, {Smith}, {Baker}, {Crifo}, {Damerdji}, {David}, {Dolding},
  {Fr{\'e}mat}, {Gosset}, {Guerrier}, {Guy}, {Haigron}, {Jan{\ss}en},
  {Marchal}, {Plum}, {Soubiran}, {Th{\'e}venin}, {Ajaj}, {Allende Prieto},
  {Babusiaux}, {Boudreault}, {Chemin}, {Delle Luche}, {Fabre}, {Gueguen},
  {Hambly}, {Lasne}, {Meynadier}, {Pailler}, {Panem}, {Royer}, {Tauran},
  {Zurbach}, {Zwitter}, {Arenou}, {Bossini}, {Gerssen}, {G{\'o}mez},
  {Lemaitre}, {Leclerc}, {Morel}, {Munari}, {Turon}, {Vallenari}, \& {{\v
  Z}erjal}}]{2019A&A...622A.205K}
{Katz}, D., {Sartoretti}, P., {Cropper}, M., {et~al.} 2019, \aap, 622, A205,
  \dodoi{10.1051/0004-6361/201833273}

\bibitem[{{Kordopatis} {et~al.}(2013){Kordopatis}, {Gilmore}, {Steinmetz},
  {Boeche}, \& others.}]{2013AJ....146..134K}
{Kordopatis}, G., {Gilmore}, G., {Steinmetz}, M., {Boeche}, C., \& others.
  2013, \aj, 146, 134, \dodoi{10.1088/0004-6256/146/5/134}

\bibitem[{{Kunder} {et~al.}(2017){Kunder}, {Kordopatis}, {Steinmetz},
  {Zwitter}, \& others.}]{2017AJ....153...75K}
{Kunder}, A., {Kordopatis}, G., {Steinmetz}, M., {Zwitter}, T., \& others.
  2017, \aj, 153, 75, \dodoi{10.3847/1538-3881/153/2/75}

\bibitem[{{Lee} {et~al.}(2008){Lee}, {Beers}, {Sivarani}, {Allende Prieto},
  {Koesterke}, {Wilhelm}, {Re Fiorentin}, {Bailer-Jones}, {Norris}, {Rockosi},
  {Yanny}, {Newberg}, {Covey}, {Zhang}, \& {Luo}}]{2008AJ....136.2022L}
{Lee}, Y.~S., {Beers}, T.~C., {Sivarani}, T., {et~al.} 2008, \aj, 136, 2022,
  \dodoi{10.1088/0004-6256/136/5/2022}

\bibitem[{{Liu} {et~al.}(2019){Liu}, {Fu}, {Zong}, {Shi}, {Luo}, {Zhang},
  {Cui}, {Hou}, {Pan}, {Shan}, {Chen}, {Bai}, {Chen}, {Du}, {Hou}, {Liu},
  {Tian}, {Wang}, {Wang}, {Wu}, {Wu}, {Yan}, \& {Zuo}}]{2019arXiv190100619L}
{Liu}, N., {Fu}, J.-N., {Zong}, W., {et~al.} 2019, arXiv e-prints.
\newblock \doarXiv{1901.00619}

\bibitem[{{Luo} {et~al.}(2015){Luo}, {Zhao}, {Zhao}, {Deng}, \&
  others.}]{2015RAA....15.1095L}
{Luo}, A.-L., {Zhao}, Y.-H., {Zhao}, G., {Deng}, L.-C., \& others. 2015,
  Research in Astronomy and Astrophysics, 15, 1095,
  \dodoi{10.1088/1674-4527/15/8/002}

\bibitem[{{Majewski} {et~al.}(2017){Majewski}, {Schiavon}, {Frinchaboy}, \&
  others.}]{2017AJ....154...94M}
{Majewski}, S.~R., {Schiavon}, R.~P., {Frinchaboy}, P.~M., \& others. 2017, AJ,
  154, 94, \dodoi{10.3847/1538-3881/aa784d}

\bibitem[{{Martin} {et~al.}(2019){Martin}, {Triaud}, {Udry}, {Marmier},
  {Maxted}, {Collier Cameron}, {Hellier}, {Pepe}, {Pollacco}, {Segransan}, \&
  {West}}]{2019arXiv190101627M}
{Martin}, D.~V., {Triaud}, A.~H.~M.~J., {Udry}, S., {et~al.} 2019, arXiv
  e-prints.
\newblock \doarXiv{1901.01627}

\bibitem[{McKinney(2011)}]{pythonpandas}
McKinney, W. 2011, pandas : powerful Python data analysis toolkit.
\newblock \url{http://pandas.sourceforge.net/}

\bibitem[{Oliphant(2006)}]{oliphant_guide_2006}
Oliphant, T. 2006, Guide to {NumPy} (Trelgol Publishing).
\newblock \url{http://www.tramy.us/numpybook.pdf}

\bibitem[{{Ricker} {et~al.}(2014){Ricker}, {Winn}, {Vanderspek}, {Latham},
  {Bakos}, {Bean}, {Berta-Thompson}, {Brown}, {Buchhave}, {Butler}, {Butler},
  {Chaplin}, {Charbonneau}, {Christensen-Dalsgaard}, {Clampin}, {Deming},
  {Doty}, {De Lee}, {Dressing}, {Dunham}, {Endl}, {Fressin}, {Ge}, {Henning},
  {Holman}, {Howard}, {Ida}, {Jenkins}, {Jernigan}, {Johnson}, {Kaltenegger},
  {Kawai}, {Kjeldsen}, {Laughlin}, {Levine}, {Lin}, {Lissauer}, {MacQueen},
  {Marcy}, {McCullough}, {Morton}, {Narita}, {Paegert}, {Palle}, {Pepe},
  {Pepper}, {Quirrenbach}, {Rinehart}, {Sasselov}, {Sato}, {Seager},
  {Sozzetti}, {Stassun}, {Sullivan}, {Szentgyorgyi}, {Torres}, {Udry}, \&
  {Villasenor}}]{2014SPIE.9143E..20R}
{Ricker}, G.~R., {Winn}, J.~N., {Vanderspek}, R., {et~al.} 2014, in \procspie,
  Vol. 9143, Space Telescopes and Instrumentation 2014: Optical, Infrared, and
  Millimeter Wave, 914320

\bibitem[{{Sartoretti} {et~al.}(2018){Sartoretti}, {Katz}, {Cropper},
  {Panuzzo}, \& others.}]{2018A&A...616A...6S}
{Sartoretti}, P., {Katz}, D., {Cropper}, M., {Panuzzo}, P., \& others. 2018,
  \aap, 616, A6, \dodoi{10.1051/0004-6361/201832836}

\bibitem[{{Siebert} {et~al.}(2011){Siebert}, {Williams}, {Siviero}, \&
  others.}]{2011AJ....141..187S}
{Siebert}, A., {Williams}, M.~E.~K., {Siviero}, A., \& others. 2011, \aj, 141,
  187, \dodoi{10.1088/0004-6256/141/6/187}

\bibitem[{{Soubiran} {et~al.}(2013){Soubiran}, {Jasniewicz}, {Chemin}, {Crifo},
  {Udry}, {Hestroffer}, \& {Katz}}]{2013A&A...552A..64S}
{Soubiran}, C., {Jasniewicz}, G., {Chemin}, L., {et~al.} 2013, \aap, 552, A64,
  \dodoi{10.1051/0004-6361/201220927}

\bibitem[{{Soubiran} {et~al.}(2018){Soubiran}, {Jasniewicz}, {Chemin},
  {Zurbach}, {et~al.}}]{2018A&A...616A...7S}
{Soubiran}, C., {Jasniewicz}, G., {Chemin}, L., {Zurbach}, C., {et~al.} 2018,
  \aap, 616, A7, \dodoi{10.1051/0004-6361/201832795}

\bibitem[{{Steinmetz} {et~al.}(2006){Steinmetz}, {Zwitter}, {Siebert},
  {Watson}, \& others.}]{2006AJ....132.1645S}
{Steinmetz}, M., {Zwitter}, T., {Siebert}, A., {Watson}, F.~G., \& others.
  2006, AJ, 132, 1645, \dodoi{10.1086/506564}

\bibitem[{{Wu} {et~al.}(2011){Wu}, {Luo}, {Li}, {Shi}, {Prugniel}, {Liang},
  {Zhao}, {Zhang}, {Bai}, {Wei}, {Dong}, {Zhang}, \&
  {Chen}}]{2011RAA....11..924W}
{Wu}, Y., {Luo}, A.-L., {Li}, H.-N., {et~al.} 2011, Research in Astronomy and
  Astrophysics, 11, 924, \dodoi{10.1088/1674-4527/11/8/006}

\bibitem[{{Yanny} {et~al.}(2009){Yanny}, {Rockosi}, {Newberg}, {Knapp}, \&
  others.}]{2009AJ....137.4377Y}
{Yanny}, B., {Rockosi}, C., {Newberg}, H.~J., {Knapp}, G.~R., \& others. 2009,
  AJ, 137, 4377, \dodoi{10.1088/0004-6256/137/5/4377}

\bibitem[{Zaharia {et~al.}(2016)Zaharia, Xin, Wendell, Das, Armbrust, Dave,
  Meng, Rosen, Venkataraman, Franklin, Ghodsi, Gonzalez, Shenker, \&
  Stoica}]{Zaharia:2016:ASU:3013530.2934664}
Zaharia, M., Xin, R.~S., Wendell, P., {et~al.} 2016, Commun. ACM, 59, 56,
  \dodoi{10.1145/2934664}

\bibitem[{{Zhao} {et~al.}(2012){Zhao}, {Zhao}, {Chu}, {Jing}, \&
  {Deng}}]{2012RAA....12..723Z}
{Zhao}, G., {Zhao}, Y.-H., {Chu}, Y.-Q., {Jing}, Y.-P., \& {Deng}, L.-C. 2012,
  Research in Astronomy and Astrophysics, 12, 723,
  \dodoi{10.1088/1674-4527/12/7/002}

\bibitem[{{Zwitter} {et~al.}(2008){Zwitter}, {Siebert}, {Munari}, {Freeman}, \&
  others.}]{2008AJ....136..421Z}
{Zwitter}, T., {Siebert}, A., {Munari}, U., {Freeman}, K.~C., \& others. 2008,
  \aj, 136, 421, \dodoi{10.1088/0004-6256/136/1/421}

\end{thebibliography}

\end{document}